\documentclass[aps,prd,preprintnumbers,superscriptaddress]{revtex4}
\usepackage{amsmath}
\usepackage{amssymb}
\usepackage{amsthm}
\usepackage{todonotes}
\usepackage{mathtools}
\usepackage{mathrsfs}
\usepackage{bm}
\usepackage{slashed} 
\usepackage{graphicx}
\usepackage{multirow}
\usepackage{tikz}
\usepackage[caption=false]{subfig}
\usepackage{relsize}	
\usepackage{array}
\usepackage{float}
\usepackage{color}
\usepackage{xcolor}
\usepackage{soul}
\usepackage{verbatim} 
\usepackage{hyperref}
\usepackage{pifont}
\newcommand{\cmark}{\ding{51}}%
\newcommand{\xmark}{\ding{55}}
\allowdisplaybreaks
\newcommand{\beq}{\begin{equation}}
	\newcommand{\eeq}{\end{equation}}
\newcommand{\be}{\begin{eqnarray}}
	\newcommand{\ee}{\end{eqnarray}}

\newcommand{\bsq}{{\boldsymbol q}^{\perp}}

\newcommand{\bsk}{{\boldsymbol k}^{\perp}}
\newcommand{\bska}{{\boldsymbol\kappa}^\perp}
\newcommand{\bskapr}{{\boldsymbol\kappa}^{\prime \perp}}
\newcommand{\bskasq}{{\kappa}^{\perp2}}

\newcommand{\bsb}{{\boldsymbol b}^{\perp}}

\newcommand{\bsP}{{\boldsymbol P}^{\perp}} 
 
\newcommand{\bsp}{{\boldsymbol p}^{\perp}} 
\newcommand{\bspp}{{\boldsymbol p}^{\perp 2}}

\newcommand{\es}{&=&}

\newcommand{\bsD}{\boldsymbol{\Delta}^{\perp}}

\newcommand{\bs}{\boldsymbol}
\begin{document}
	
	\begin{titlepage}
		
		\title{Spatial distribution of Angular Momentum Inside  a Quark  State Dressed with a  Gluon }
		
		\author{Ravi Singh}
		\email{30004559@iitb.ac.in}
		\author{Sudeep Saha}
		\email{sudeepsaha@iitb.ac.in} 
		\author{Asmita Mukherjee}
		\email{asmita@phy.iitb.ac.in}
		\affiliation{Department of Physics,
			Indian Institute of Technology Bombay, Powai, Mumbai 400076,
			India}
		\author{Nilmani Mathur}
		\email{nilmani@theory.tifr.res.in}
		\affiliation{ Department of Theoretical Physics,
			Tata Institute of Fundamental Research, Colaba, Mumbai 400005,
			India}
		\begin{abstract}
			{{We investigate the different decompositions of the angular momentum in QCD for a relativistic spin $1/2$ composite state, namely a quark dressed with a gluon. We use light-front Hamiltonian perturbation theory, and in the light-front gauge, use the two-component framework by eliminating the constrained degrees of freedom. We also investigate the different decompositions of the angular momentum at the level of two-dimensional densities in the front form, including the effect of the so-called potential term. In this work, we consider the contribution coming from the quark part of the energy-momentum tensor. We contrast the different decompositions and also compare with other calculations in the literature. We also present the gravitational form factor related to the antisymmetric part of the energy-momentum tensor. }}  
		\end{abstract}
		\date{\today}
		\maketitle
		
	\end{titlepage}
	\section{Introduction}
	
	{{Experimental results have established that only one-third of the proton spin comes from the quark's intrinsic spin \cite{EuropeanMuon:1987isl, EuropeanMuon:1989yki, Kuhn:2008sy, Anselmino:1994gn}}}. Recently, RHIC-spin experiments have provided important constraints on the contribution of gluon's helicity to the proton spin \cite{deFlorian:2014yva, Ji:2013fga}. 
	Since the contributions from
	the intrinsic spin of quarks and gluons do not account for the total nucleon spin, orbital angular momentum (OAM)\raisebox{0.3ex}{\textsuperscript{1}} \footnotetext[1]{Assuming the Jaffe-Manohar decomposition in light-cone gauge, see section-II.}{is a good candidate }. 
	{It is anticipated that future experiments} like JLab 12 GeV \cite{Dudek:2012vr} and Electron-Ion Collider (EIC) at BNL \cite{AbdulKhalek:2021gbh} will provide high-precision measurements of orbital angular momentum contribution of quarks and gluons respectively\cite{Ji:2020ena}. {{{On the theory hand, the decomposition of the nucleon spin into intrinsic components and OAM of its constituent quarks and gluons has been discussed extensively in the literature in the past few years \cite{Leader:2013jra}}. The main issue for a long time was the fact that the total spin, in particular, the angular momentum of the gluons cannot be separated further into intrinsic and orbital parts in a gauge-invariant manner. {On the contrary}, polarized electron-proton and proton-proton scattering experiments have since measured spin asymmetries that are sensitive to gluon helicity distribution, {or to the intrinsic spin} of the gluons in the nucleon. {Hence, one needs to identify a gauge invariant observable with the one that has been probed in the experiments}. It has later been shown that indeed, a gauge-invariant decomposition of gluon angular momentum is possible. The procedure used to achieve this gauge-invariant decomposition as shown in \cite{Chen:2008ag, Chen:2009mr, Wakamatsu:2010qj, Wakamatsu:2010cb} adds another term to the decomposition called the ``potential angular momentum" \cite{Burkardt:2012sd}. The potential angular momentum, being gauge-invariant in itself, can be added to  the OAM of the quark or the OAM of the gluon; thus there is an ambiguity even in the definition of the total angular momentum of quarks/gluons. In fact, infinitely many decompositions of the nucleon spin can be theoretically possible, although only a few of them give meaningful physical insight into the spin structure of nucleons namely, Belinfante \cite{Belinfante1939OnTS}, Ji \cite{Ji_1997}, Jaffe-Manohar \cite{Jaffe:1989jz}, Chen et. al. \cite{Chen:2008ag, Chen:2009mr} and Wakamatsu decomposition \cite{Wakamatsu:2010qj}. A comprehensive review of all these decompositions, the theoretical issues in this line, and how they are addressed can be {found in Ref.} \cite{Leader:2013jra} and we will also discuss them briefly in section II.


			The different angular momentum decompositions give the same expression for total angular momentum at the integrated level but not at the density level \cite{PhysRevD.94.114021}.
			This is because these decompositions differ from each other by superpotential (surface) terms that vanish after integration \cite{Leader:2013jra}.
			{{So, the total angular momentum density cannot be interpreted as a sum of the OAM and spin density only and it becomes}}
			{{interesting to investigate the spin and OAM of the quarks and gluons at a density level and explore how the different decompositions affect these densities. As shown by Burkardt \cite{Burkardt:2002hr, Burkardt:2000wc}, generalized parton distributions (GPDs) contain information about the spatial distribution of quarks and gluons, as the Fourier transformation of the GPDs gives the parton distributions in the transverse impact parameter space, which have density interpretation in the frame where the momentum transfer is purely in the transverse direction. GPDs can be extracted from {the exclusive processes like} deeply virtual Compton scattering (DVCS) and deeply virtual meson production (DVMP)\cite{diehl2003generalized, Boffi:2007yc, Bacchetta:2016ccz, dHose:2016mda, Kumericki:2016ehc, Collins:1996fb}.  The gravitational form factors, which are related to the GPDs through a moment, give the pressure, shear, and energy distributions due to the quarks and gluons in the nucleon \cite{Burkert:2018bqq, Burkert:2021ith, Lorce:2018egm, Polyakov:2018zvc}. Similarly, the angular momentum distributions inside the nucleon can be obtained \cite{Lorce:2017wkb}. Such distributions in the literature are usually constructed {in the following two different approaches}; in the Breit frame, one calculates the 3D distributions \cite{Polyakov:2002yz, Goeke:2007fp}, however, such distributions need relativistic corrections unless the nucleon is considered to be infinitely massive. On the other hand, one can construct 2D distribution in the light-front framework. Due to the transverse Galilean symmetry on the light front, these are free from relativistic corrections.}}
			{{The presence of surface terms that connect various decompositions affects the relationship between the Fourier transform of the GPDs in the impact parameter space and the angular momentum distributions in the transverse plane.}}
			{{Different definitions of the angular momentum distributions were investigated in Ref \cite{PhysRevD.94.114021, Lorce:2017wkb}} using a scalar-diquark model (SDQM), and it was shown that when all the surface terms are included, the total angular momentum distribution is the same for Belinfante and Ji decomposition. Distributions of quark angular momentum in a light-front quark-diquark model using soft wall AdS/QCD have been studied in \cite{Kumar:2017dbf}. The role of surface terms in the nucleon spin decomposition problem at integrated and density level has also been discussed in \cite{Wakamatsu:2019ain}. 
				The term responsible for the difference between Ji and Jaffe-Manohar decompositions of quark and gluon OAM is the potential angular momentum}\cite{Burkardt:2012sd, Hatta:2011ku}. This term gives a non-vanishing contribution to the nucleon's spin as suggested by lattice QCD studies \cite{Engelhardt:2017miy, Engelhardt:2020qtg}. {{A numerical analysis of the renormalization scale dependence of potential angular momentum in Ref. \cite{Hatta:2019csj} also points to the non-zero contribution of the potential AM}}. {{In Ref. \cite{Lorce:2017wkb}, it was shown that}}  the Ji and Jaffe-Manohar definitions of angular momentum coincide in SDQM, as expected in a system without 
			gauge bosons, {{due to the vanishing of the potential angular momentum.}}
			This observation has also been confirmed {more recently in Ref.} \cite{Amor-Quiroz:2020qmw} at the two loop-level in SDQM.
			{{The authors}} proposed that since potential AM is related to the torque acting on the quark, {just the sole presence of a gauge boson is not enough for its measurement}.
			In a two-body system, there cannot be classically any Lorentz torque exerted by the spectator on the struck constituent despite the presence of a Lorentz force.
			{{The potential term is also found to be zero for an electron in QED  at one loop level \cite{Ji:2015sio}.}}
			
			In this work, we investigate the different decompositions of the angular momentum and their 
			contribution to the intrinsic spin and orbital angular momentum distributions. Following the approach of 
			\cite{Lorce:2017wkb} we calculate the spatial distributions on the light front using overlaps in terms of the light-front wave functions in light-front Hamiltonian perturbation theory. In this analysis, instead of a nucleon state in which it is highly non-trivial to include gluons, we consider a relativistic spin-1/2 composite state of a quark dressed with a gluon at one-loop in QCD. In comparison to a nucleon, it is a simple state with a gluonic degree of freedom \cite{Harindranath:1998pd, harindranath1996hamiltonian}. We have recently used such a state to investigate the pressure and shear distributions \cite{More:2021stk, More:2023pcy} related to the gravitational form factors.  
			The advantage is that the LFWFs for such a state can be calculated analytically using the light-front QCD Hamiltonian and it incorporates the full quark-gluon interactions up to one loop \cite{Harindranath:2001rc}.
			These LFWFs are boost-invariant because they can be written in terms of relative momenta which are frame-independent \cite{Brodsky:1997de}. We use the two-component formalism developed in  \cite{Zhang:1993dd}, where in the light-front gauge the constrained degrees of freedom are removed using equations of constraint.  This allows for an analytical calculation of matrix elements for all components of the energy-momentum tensor (EMT) relevant to the angular momentum distributions. 
			Also, the transverse boost invariance in light-front dynamics makes it possible to separate the dynamics associated with the centre of mass and the internal dynamics in the calculation of the longitudinal component of angular momentum \cite{Harindranath:1998ve}.}} 
	{Thus, using the LFWFs of the dressed quark state, we analyze different decompositions of angular momentum at the density level, terms responsible for the difference between them, namely, superpotential and potential AM. We also investigate the observables corresponding to these distributions, like, the form factors of the EMT called the gravitational form factor (GFF).} 
	{{The $D$ form factor is related to the}} {{antisymmetric}} {{part of the EMT: this corresponds to the spin density. In this work, we investigate the contributions to the angular momentum distributions coming from the quark part of the EMT, explicitly taking into account the quark-gluon interaction. Contribution coming from the gluon part of the EMT will be presented in a separate publication.}} 
	
	{{The paper is arranged in the following manner: in section II, we discuss the different decompositions of angular momentum density in the literature; in section III we describe the two-component formalism in the light-front Hamiltonian approach, in section IV we present the angular momentum densities in the front form, in section V we present the extraction of the D form factor. Numerical results are given in section VI and conclusion in section VII.}}

	\section{Different decompositions of angular momentum density}
	
	We start off by writing the generalized form of angular momentum tensor
	\begin{align}
		J^{\mu\nu\rho}= x^{\nu}T^{\mu\rho}-x^{\rho}T^{\mu\nu}+ S^{\mu\nu\rho} = L^{\mu\nu\rho} + S^{\mu\nu\rho}, 
	\end{align}
	where $T^{\mu\nu}$ is the energy-momentum tensor (EMT) and the second term is the intrinsic part of the total angular momentum.
	{One can get multiple} decompositions of $T^{\mu \nu}$ and $J^{\mu \nu \rho}$ from the QCD Lagrangian 
	\begin{align}
		\mathcal{L}_{\text{QCD}} = \overline{\psi} \left(\frac{i}{2} \gamma_\mu \overleftrightarrow{\partial}^\mu - m \right) \psi + g \overline{\psi}\gamma_\mu A^\mu \psi - \frac{1}{2}\text{Tr}\left[G^{\mu \nu} G_{\mu \nu}\right],
	\end{align}
	where  $\overleftrightarrow{\partial}^{\mu}=\overrightarrow{\partial}^{\mu}-\overleftarrow{\partial}^{\mu}$ and the field strength is $G^{\mu\nu}(x)=\partial^{\mu}A_a^{\nu}(x)-\partial^{\nu}A_a^{\mu}(x)-ig\left[A_a^{\mu}(x),A_b^{\nu}(x)\right]$.
	The canonical EMT and generalized angular momentum densities are derived through the application of Noether's theorem, which establishes a connection between conserved currents and global space-time symmetries. They are given as
	\begin{align}
		T^{\mu \nu}(x) &= \frac{1}{2} \overline{\psi} (x) \gamma^\mu i\overleftrightarrow{\partial^\nu} \psi (x) -2\text{Tr}\left[ G^{\mu \alpha } \partial^\nu A_\alpha \right] - g^{\mu \nu} \mathcal{L}_{\text{QCD}}, \label{T_can_full} \\
		J^{\mu \nu \rho}(x)&= \frac{1}{2} \overline{\psi} (x) \gamma^\mu x^{[\nu} i \overleftrightarrow{\partial}^{\rho]} \psi(x) + \frac{1}{2}\epsilon^{\mu\nu\rho\sigma}\overline{\psi}(x)\gamma_{\sigma}\gamma_{5}\psi(x) - 2\text{Tr}\left[ G^{\mu [ \nu} A^{\rho]} \right] - 2\text{Tr}\left[ G^{\mu \alpha} x^{[\nu}\partial^{\rho]} A_\alpha \right] -x^{[\nu}g^{\rho]\mu}\mathcal{L}_{\text{QCD}}.  \label{S_can_full}
	\end{align} 
	This is referred to as the Jaffe-Manohar (JM) decomposition \cite{Jaffe:1989jz}.
	In this decomposition, the total angular momentum density of both quark and gluon is separable into orbital and intrinsic parts. {{As stated in the introduction, in this work, we calculate the quark contribution to the angular momentum densities. The analysis of gluon angular momentum densities will be a subject of future publication.}} Thus Eq. \eqref{T_can_full},\eqref{S_can_full} reduces to
	\begin{align}
		T^{\mu \nu}_\text{q}(x) &= \frac{1}{2} \overline{\psi} (x) \gamma^\mu i\overleftrightarrow{\partial^\nu} \psi (x), \label{T_can} \\
		J^{\mu \nu \rho}_\text{q}(x)&=L^{\mu \nu \rho}_\text{q} (x)+S^{\mu \nu \rho}_\text{q} (x) , \nonumber \\
		&= \frac{1}{2} \overline{\psi} (x) \gamma^\mu x^{[\nu} i \overleftrightarrow{\partial}^{\rho]} \psi(x) + \frac{1}{2}\epsilon^{\mu\nu\rho\sigma}\overline{\psi}(x)\gamma_{\sigma}\gamma_{5}\psi(x).  \label{S_can}
	\end{align} 
	
	$T^{\mu \nu}_q$ is neither gauge invariant nor symmetric. 
	So a superpotential term (a total divergence term) is added to both EMT and generalized angular momentum tensors to form Belinfante-improved tensors \cite{Belinfante1939OnTS, BELINFANTE1940449, Rosenfeld}.
	We write the Belinfante-improved form as
	\begin{align}
		T^{\mu\nu}_{\text{Bel,q}}(x)=&T_{\text{q}}^{\mu\nu}(x)+\partial_{\alpha}G^{\alpha\mu\nu}(x), \nonumber \\
		=& \frac{1}{4}\overline{\psi}(x)\left[\gamma^{\mu}i\overleftrightarrow{D}^{\nu}+\gamma^{\nu}i\overleftrightarrow{D}^{\mu}\right]\psi(x) ,\label{T_Bel} \\
		J^{\mu\nu\rho}_{\text{Bel,q}}(x)=&J_{\text{q}}^{\mu\nu\rho}(x)+\partial_{\sigma}\left[x^{\nu}G^{\sigma\mu\rho}(x) - x^{\rho}G^{\sigma\mu\nu}(x)\right], \nonumber \\
		=& \frac{1}{4}\overline{\psi}\gamma^\mu x^{[\nu}i\overleftrightarrow{D}^{\rho]} \psi + \frac{1}{4}x^{[\nu}\overline{\psi}\gamma^{\rho]} i \overleftrightarrow{D}^\mu \psi, 
	\end{align}
	where $D^\mu = \partial^\mu - igA^\mu$ is the covariant derivative acting on the quark fields. The superpotential is given by
	\begin{align}
		G^{\mu\nu\rho} = \frac{1}{4} \epsilon^{\mu\nu\rho\sigma}\overline{\psi}(x)\gamma_{\sigma}\gamma_{5}\psi(x),
	\end{align}
	and
	\begin{align}
		G^{\mu\nu\rho}(x)=\frac{1}{2}\left[S_\text{q}^{\mu\nu\rho}(x)+S_\text{q}^{\nu\rho\mu}(x)+S_\text{q}^{\rho\nu\mu}(x)\right]=-G^{\nu\mu\rho}(x).
	\end{align}
	It is antisymmetric under the exchange of the first two indices. The Belinfante-improved tensors are gauge invariant and symmetric and it can be seen that 
	\begin{align}
		J^{\mu\nu\rho}_{\text{Bel,q}}(x)=x^{\nu}T^{\mu\rho}_{\text{Bel,q}}(x)-x^{\rho}T^{\mu\nu}_{\text{Bel,q}}(x),
	\end{align}
	i.e., the Belinfante-improved total angular momentum density has the form of orbital angular momentum and it is not separated into spin and orbital components.
	Whereas, in the case of canonical total AM, it can be shown from the conservation of $T^{\mu\nu}$ and $J^{\mu\nu\rho}$ that
	\begin{align}
		T^{\nu \rho} - T^{\rho \nu} = - \partial_\mu S^{\mu \nu \rho},
	\end{align}
	i.e. the antisymmetric part of the EMT is related to the spin density.Although the Belinfante-improvement method gives a gauge-invariant and symmetric EMT from the canonical EMT, it involves adding a superpotential term to the canonical EMT in an ad-hoc manner. A standard way to obtain a symmetric and gauge-invariant Belinfante EMT is through functional variation of an action of QCD coupled to a weak external gravitational field with respect to the metric \cite{Polyakov:2018zvc, Blaschke:2016ohs}. Another rigorous method has been proposed in Ref. \cite{Freese:2021jqs} to obtain an expression of symmetric and gauge-invariant EMT by taking into account the symmetry of action under local space-time translations instead of global space-time translations.
	
	Another decomposition that was proposed by Ji \cite{Ji_1997}, also known as kinetic EMT \cite{Leader:2013jra, Lorce:2015lna}, is given as 
	\begin{align}
		T^{\mu\nu}_{\text{kin,q}}(x)=& \frac{1}{2}\overline{\psi}(x)\gamma^{\mu}i\overleftrightarrow{D}^{\nu}\psi(x). \label{T_kin} 
	\end{align}
	Unlike Jaffe-Manohar's decomposition, it is gauge-invariant, but it is also asymmetric and thus contributes to the quark's spin density. The kinetic generalized angular momentum tensor reads 
	\begin{align}
		J^{\mu\nu\rho}_{\text{kin,q}}(x)=L^{\mu\nu\rho}_{\text{kin,q}}(x)+S_{\text{q}}^{\mu\nu\rho}(x), \label{Ji_graph}
	\end{align}
	with 
	\begin{align}
		L^{\mu\nu\rho}_{\text{kin,q}}(x)=&\frac{1}{2}\overline{\psi}\gamma^\mu x^{[\nu}i\overleftrightarrow{D}^{\rho]} \psi, \\
		S_{\text{q}}^{\mu\nu\rho}(x)=&\frac{1}{2}\epsilon^{\mu\nu\rho\sigma}\overline{\psi}(x)\gamma_{\sigma}\gamma_{5}\psi(x) \label{S_kin}.
	\end{align}
	The kinetic and Belinfante-improved tensors in QCD are related as \cite{Lorce:2017wkb}
	\begin{align}
		T^{\mu\nu}_{\text{kin,q}}(x)=&T^{\mu\nu}_{\text{Bel,q}}(x)-\frac{1}{2}\partial_{\alpha}S_{\text{q}}^{\alpha\mu\nu}(x), \label{superpotential} \\
		L^{\mu\nu\rho}_{\text{kin,q}}(x)+S_{\text{q}}^{\mu\nu\rho}(x)=& J^{\mu\nu\rho}_{\text{Bel,q}}(x)-\frac{1}{2}\partial_{\sigma}\left[x^{\nu}S_{\text{q}}^{\sigma\mu\rho}(x)-x^{\rho}S_{\text{q}}^{\sigma\mu\nu}(x)\right].
	\end{align}
	
	It is clear that the canonical, Belinfante-improved, and kinetic tensors give the same conserved charges since they differ by the total derivative of the superpotential.
	So at the total integration level, they give the same results. Still, at the level of densities or distributions, where the total divergence {terms do not vanish}, these decompositions give different results.
	
	As previously mentioned, in the JM decomposition, the total angular momentum of the gluon is split into spin and orbital angular momentum. Thus, the decomposition is complete; but except for the quark spin, all the other contributions are gauge non-invariant. This is remedied by the Chen et. al. decomposition \cite{Chen:2008ag, Chen:2009mr}. It is done by separating the gauge field into two parts, longitudinal and transverse. The former is the pure-gauge part which is related to gauge symmetry and unphysical gauge degrees of freedom and the latter is related to the physical degrees of freedom, namely the two physical polarizations.
	\begin{align}
		\boldsymbol{A}= \boldsymbol{A}_{\text{\text{pure}}}+\boldsymbol{A}_{\text{\text{phys}}}.
	\end{align}
	These terms are constrained as follows 
	\begin{align}
		\boldsymbol{\nabla}\times \boldsymbol{A}_{\text{\text{pure}}}= \boldsymbol{0}, ~~~ \boldsymbol{\nabla}\cdot \boldsymbol{A}_{\text{\text{phys}}}=0.
	\end{align}
	These constraints imply that $\boldsymbol{A}_{\text{\text{pure}}}=-\boldsymbol{\nabla}\alpha_{\text{\text{pure}}}$, where $\alpha_{\text{\text{pure}}}$ is some scalar function. Also the magnetic field $\boldsymbol{B}=\boldsymbol{\nabla}\times \boldsymbol{A}_{\text{\text{phys}}}$.
	Although gauge invariant, the fields involved are non-local, for e.g.,
	\begin{align}
		\boldsymbol{A}_{\text{\text{phys}}}=\boldsymbol{A}- \boldsymbol{\nabla}\frac{1}{\boldsymbol{\nabla}^2}\boldsymbol{\nabla}\cdot \boldsymbol{A}, && \frac{1}{\boldsymbol{\nabla}^2}f(\boldsymbol{x})=-\frac{1}{4\pi}\int d^3x^{\prime}\frac{f(\boldsymbol{x}^{\prime})}{|\boldsymbol{x}-\boldsymbol{x}^{\prime}|}, 
	\end{align}
	where $\frac{1}{\boldsymbol{\nabla}^2}$ is an integral operator. 
	This non-locality vanishes if one works in the Coulomb gauge, $\boldsymbol{\nabla \cdot A }= 0 $ or in any other gauge where $ A = A_{\text{\text{phys}}} \text{ and } A_{\text{pure}} = 0 $. 
	With this gauge-fixing, this decomposition coincides with the JM decomposition. 
	Another issue with this decomposition is that it is defined in a specific Lorentz frame.
	Therefore, for our analysis, we consider the covariant version of Chen. et al. decomposition that was given by Wakamatsu \cite{Wakamatsu:2010cb, Wakamatsu:2010qj}.
	This is called gauge-invariant canonical (gic) decomposition. The potential angular momentum term is added to the quark OAM part in this {gauge-invariant extension and one arrives} 
	\begin{align}
		T^{\mu \nu}_{\text{gic,q}}(x) &= \frac{1}{2} \overline{\psi} (x) \gamma^\mu i\overleftrightarrow{D}^\nu_{\text{pure}} \psi (x), 
		\label{T_gic} \\
		J^{\mu \nu \rho}_{\text{gic,q}}(x)&=\frac{1}{2} \overline{\psi} (x) \gamma^\mu x^{[\nu} i \overleftrightarrow{D}^{\rho]}_{\text{pure}} \psi(x) + \frac{1}{2}\epsilon^{\mu\nu\rho\sigma}\overline{\psi}(x)\gamma_{\sigma}\gamma_{5}\psi(x),
		\label{J_gic}
	\end{align} 
	where $D_\mu^{\text{pure}} = \partial_\mu - igA_\mu^{\text{pure}}$ for quark fields. The EMT and the AM densities of canonical and gauge-invariant canonical decomposition differ by a superpotential term which vanishes in a suitable gauge
	\begin{align}
		T^{\mu\nu}(x) = T^{\mu\nu}_{\text{gic}}(x) - 2\partial_\alpha \text{Tr} \left[ G^{\mu \alpha}(x) A^\nu_{\text{pure}}(x) \right], \\
		J^{\mu \nu \rho}(x) = J^{\mu \nu \rho}_{\text{gic}}(x) - 2\partial_\alpha \text{Tr} \left[ G^{\mu\alpha}(x) x^{[\nu}A^{\rho]}_{\text{pure}} (x) \right].
	\end{align}
	
	Wakamatsu also proposed another gauge-invariant decomposition which is similar to Ji's decomposition with the advantage that the gluon total angular momentum is split into spin and orbital parts. The potential angular momentum is associated with the gluon OAM part in this decomposition. Hence, the quark part is exactly the same as Ji's decomposition. It is called as the gauge-invariant kinetic (gik) decomposition and is given as 
	\begin{align}
		T^{\mu \nu }_{\text{gik,q}}(x) &= \frac{1}{2} \overline{\psi} (x) \gamma^\mu i\overleftrightarrow{D}^\nu \psi (x),
		\label{T_gik} \\
		J^{\mu \nu \rho}_{\text{gik,q}}(x) &= \frac{1}{2} \overline{\psi} (x) \gamma^\mu x^{[\nu} i \overleftrightarrow{D}^{\rho]} \psi(x) + \frac{1}{2}\epsilon^{\mu\nu\rho\sigma}\overline{\psi}(x)\gamma_{\sigma}\gamma_{5}\psi(x).
		\label{J_gik}
	\end{align}
	
	To summarise what we discussed so far in this section; the decompositions can be divided into two categories, kinetic and canonical \cite{Leader:2013jra}. The kinetic class includes Belinfante \cite{Belinfante1939OnTS}, Ji \cite{Ji_1997}, and Wakamatsu decomposition \cite{Wakamatsu:2010qj}, which is a gauge-invariant extension (GIE) of Ji decomposition. The canonical class includes Jaffe-Manohar \cite{Jaffe:1989jz} and its GIE, Chen et. al. decomposition \cite{Chen:2008ag, Chen:2009mr}.
	This classification became possible because of Wakamatsu's covariant generalization of the decomposition of QCD angular momentum tensor into five separately gauge-invariant terms \cite{Wakamatsu:2010cb, Wakamatsu:2010qj}.
	These terms are spin and orbital angular momentum (OAM) of quarks and gluons and potential angular momentum.
	The potential angular momentum, being gauge-invariant in itself, can be added to quark's OAM or gluon's OAM giving canonical and kinetic families respectively. 
	An important point to note about the various angular momentum decompositions is that the individual components do not always satisfy the angular momentum commutation relations, $\left[J^i,J^j\right] = i \epsilon^{ijk} J^k$ \cite{Leader:2013jra, LEADER201923}. 
	For instance, none of the terms in the Belinfante decomposition individually follow the commutation relations, rendering them unsuitable for direct interpretation as generators of rotation.
	In Jaffe-Manohar and Chen et al. decompositions, quark OAM and spin both follow the angular momentum commutation relations individually. 
	The gluon OAM and spin cannot be individually interpreted as generators of rotation  \cite{vanEnk1994497, doi:10.1080/09500341003654427}. 
	In Ji and Wakamatsu's decomposition, only the quark spin can be considered as a generator of rotations as it follows the SU(2) algebra. Following is the table that lists the properties of all the angular momentum densities derived from EMTs of kinetic and canonical family.\raisebox{0.3ex}{\textsuperscript{2}} \footnotetext[2]{Since the gluon OAM and spin term are not derived in this work, one can verify their angular momentum commutation property using the Ref. \cite{Leader:2013jra, LEADER201923}.}

	
	\begin{table}[ht]
		\begin{tabular}{|c|c|c|c|c|}
			\hline
			\textbf{Class}  & \textbf{EMT}         & \textbf{AM densities} & \textbf{Gauge invariant} & \textbf{Follow $SU(2)$ algebra} \\ \hline
			\multirow{9}{*}{\textbf{Kinetic}}   & \multirow{2}{*}{\textbf{Belinfante}}    &         $\boldsymbol{J_\text{Bel,q}}$             &       \cmark                   &      \xmark                         \\ \cline{3-5} 
			&                                         &              $\boldsymbol{J_\text{Bel,g}}$             &                        \cmark  &         \xmark                        \\ \cline{2-5} 
			& \multirow{3}{*}{\textbf{Ji}}            &          $\boldsymbol{L_\text{Ji,q}}$              &       \cmark                   &   \xmark                              \\ \cline{3-5} 
			&                                         &            $\boldsymbol{S_\text{Ji,q}} $              &   \cmark                       &     \cmark                            \\ \cline{3-5} 
			&                                         &             $\boldsymbol{J_\text{Ji,g}}$              &          \cmark                &       \xmark                          \\ \cline{2-5} 
			& \multirow{4}{*}{\textbf{Wakamatsu (gik)}}     &         $\boldsymbol{L_\text{gic,q}}$               &     \cmark                     &     \xmark                            \\ \cline{3-5} 
			&                                         &           $\boldsymbol{S_\text{gic,q}}$                &   \cmark                       &    \cmark                             \\ \cline{3-5} 
			&                                         &             $\boldsymbol{L_\text{gic,g}}$            &    \cmark                      &    \xmark                             \\ \cline{3-5} 
			&                                         &             $\boldsymbol{S_\text{gic,g}}$              &   \cmark                       &    \xmark                             \\ \hline
			\multirow{8}{*}{\textbf{Canonical}} & \multirow{4}{*}{\textbf{Jaffe-Manohar}} &         $\boldsymbol{L_\text{JM,q}}$                  &         \xmark                 &           \cmark                      \\ \cline{3-5} 
			&                                         &            $\boldsymbol{S_\text{JM,q}}$               &          \cmark                &    \cmark                            \\ \cline{3-5} 
			&                                         &         $\boldsymbol{L_\text{JM,g}}$                  &          \xmark                &               \xmark                  \\ \cline{3-5} 
			&                                         &           $\boldsymbol{S_\text{JM,g}}$                &           \xmark               &    \xmark                             \\ \cline{2-5} 
			& \multirow{4}{*}{\textbf{Chen et al. (gic)}}   &        $\boldsymbol{L_\text{gic,q}}$                &      \cmark                    &     \cmark                            \\ \cline{3-5} 
			&                                         &         $\boldsymbol{S_\text{gic,q}}$                  &          \cmark                &             \cmark                    \\ \cline{3-5} 
			&                                         &            $\boldsymbol{L_\text{gic,g}}$               &          \cmark                &    \xmark                             \\ \cline{3-5} 
			&                                         &            $\boldsymbol{S_\text{gic,g}}$               &           \cmark               &     \xmark                            \\ \hline
		\end{tabular}
		\caption{Properties of all the angular momentum densities derived from EMTs of kinetic and canonical family.}
	\end{table}
	\section{The two-component formalism in light-front Hamiltonian approach}
	{{In this section, we describe the formalism used to calculate the distributions of various constituents of angular momentum densities.}} {{We use the two-component formalism \cite{Zhang:1993dd}  of light-front Hamiltonian QCD, where in the light-front gauge $A^+=0$, one can eliminate the unphysical degrees of freedom using the equations of constraint.}} {{The quark field can be decomposed as $\psi=\psi^{+}+\psi^{-}$, $\psi_{\pm}=\Lambda_{\pm}\psi=\frac{1}{2}\gamma^{0}\gamma^{\pm}\psi$, where $\Lambda_{\pm}$ are the projection operators for the corresponding fields. The $\psi^{-}$ component and the longitudinal component of the gauge field $A^{-}$ are constrained fields and can be written in terms of $\psi^{+}$ and $\bs{A}^{\perp}$ in the following way 
			\begin{align}
				i\partial^+\psi_-=&\left(i\alpha^{\perp}\cdot \partial^{\perp}+g \alpha^{\perp}\cdot \bs{A}^{\perp}+\beta m\right)\psi_+,\\
				\frac{1}{2}\partial^+E_a^-=&\left(\partial^iE_a^i+gf^{abc}A_b^iE_c^i\right)-g\psi_+^{\dagger}T^a\psi_+,
			\end{align}
			where $T^{a}$ the Gell-Mann SU(3) matrices: $[T^{a},T^{b}]=if^{abc}T^{c}$ and $\text{Tr}(T^{a}T^{b})=\frac{1}{2}\delta_{ab}$, $m$ is the quark mass, $\alpha^{\perp}=\gamma^0\gamma^{\perp}$, $\beta=\gamma^0$ and $E_a^{-,i}=-\frac{1}{2}\partial^{+}A_a^{-,i}$, $(i=1,2)$. So in light-front QCD, the independent degrees of freedom are $\psi^{+}$ and $\bs{A}^{\perp}$. It is now possible to reduce a four-component fermion field to a two-component field with a suitable choice of the light-front representation of the gamma matrices defined by \cite{Zhang:1993dd}
			\begin{align}
				\gamma^{+}=&\begin{pmatrix}
					0 & 0 \\
					2i & 0
				\end{pmatrix} , ~~\gamma^{-}=\begin{pmatrix}
					0 & -2i \\
					0 & 0
				\end{pmatrix}\\
				\gamma^{i}=&\begin{pmatrix}
					-i\sigma^{i} & 0 \\
					0 & i\sigma^{i}
				\end{pmatrix}, ~~ \gamma^{5}=\begin{pmatrix}
					\sigma^3 & 0 \\
					0 & -\sigma^3
				\end{pmatrix}.
			\end{align} 
			In this representation, the projection operators become
			\begin{align}
				\Lambda_{+}=\begin{pmatrix}
					1 & 0 \\
					0 & 0
				\end{pmatrix}, ~~ \Lambda_{-}=\begin{pmatrix}
					0 & 0 \\
					0 & 1
				\end{pmatrix},
			\end{align}
			and the fermion field decomposes as
			\begin{align}
				\psi_+= \begin{bmatrix}
					\xi\\0
				\end{bmatrix}, ~~~~\psi_-=\begin{bmatrix}
					0\\ \eta
				\end{bmatrix}, 
			\end{align}
			where $\xi$ represents the two-component light-front quark field and $ \eta$ is constrained field: 
			\begin{align}
				\xi(y) &= \sum_{\lambda}\chi_{\lambda}\int \frac{[dk]}{\sqrt{2(2\pi)^3}}[b_{\lambda}(k)e^{-ik\cdot y}+d^{\dagger}_{-\lambda}(k)e^{ik\cdot y}], \\
				\eta(y) &= \left(\frac{1}{i\partial^+}\right)\left[\sigma^{\perp}\cdot\left(i\partial^{\perp}+g A^{\perp}(y)\right)+im\right]\xi(y),
			\end{align}
			where $[dk]= \frac{dk^+ d^2k^\perp} {\sqrt{2(2\pi)^3{k^{+}}}}$.
			The dynamical components of the gluon field are given by 
			\begin{align}
				A^{\perp}(y) = \sum_{\lambda} \int \frac{[dk]}{\sqrt{2 (2 \pi)^3 k^+}}[{\bf\epsilon}^{\perp}_{\lambda}a_{\lambda}(k)e^{-i k \cdot y}+ {\bf \epsilon}^{\perp*}_{\lambda}a^{\dagger}_{\lambda}(k)e^{i k \cdot y}],
			\end{align} 
			where $\chi_{\lambda}$ is the eigenstate of $\sigma^3$ and $\epsilon_{\lambda}^{i}$ is the polarization vector of transverse gauge field.}} {{The state can be expanded in Fock space in terms of multi-parton light-front wave functions (LFWFs).   As stated in the introduction, in this work, instead of the proton state we take a dressed quark state, that is a  quark dressed with a gluon at one loop in QCD.
			In the light-front Hamiltonian framework, we truncate the Fock space expansion up to the two-particle sector in a boost invariant way \cite{Harindranath:1998pd}. The LFWFs can be calculated analytically for such a state. 
			The dressed quark state is written as \cite{Harindranath:2001rc}
			\begin{align}
				|p,\sigma \rangle = \psi_1(p,\sigma)b^{\dagger}_{\sigma}(p)|0\rangle +\sum_{\lambda_1,\lambda_2}\int \frac{dk_1^+d^2k_1^{\perp}dk_2^{+}d^2k_2^{\perp}}{(16\pi^3)\sqrt{k_1^+k_2^+}}~ \sqrt{16\pi^3 p^+}\psi_2(p,\sigma|k_1,\lambda_1;k_2,\lambda_2)\delta^{(3)}(p-k_1-k_2)b_{\lambda_1}^{\dagger}(k_1)a_{\lambda_2}^{\dagger}(k_2)|0 \rangle. 
				\label{state}
			\end{align}
			In Eq. \ref{state}, $\psi_1(p, \sigma)$ in the first term, corresponds to a single particle with momentum (helicity) $p (\lambda)$ and also gives the normalization of the state. The two-particle LFWF, $\psi_2(p,\sigma|k_1,\lambda_1;k_2,\lambda_2)$ is related to the probability amplitude of finding two particles namely a quark and a gluon with momentum (helicity) $k_1(\lambda_1)$ and $k_2(\lambda_2)$, respectively,  inside the dressed quark state. }}
	$b^\dagger$ and $a^\dagger$ correspond to the creation operator of quark and gluon respectively. 
	
	The boost invariant light-front wavefunctions (LFWFs) can be written in terms of relative momenta so that they are independent of the momentum of the composite state \cite{Brodsky:1997de}. The relative momenta $x_i$, $\bska_i$ are defined such that they satisfy the relation $x_1+x_2=1$ and $\bska_1+\bska_2=0$.
	\be
	k_i^+=x_ip^+,~~~~ \bsk_i=\bska_i+x_i \bsp, 
	\ee 
	where $x_i$ is the longitudinal momentum fraction for the quark or gluon, inside the two-particle LFWF.
	The boost invariant two-particle LFWF can be written as, \cite{Harindranath:1998pd, Harindranath:2001rc}
	
	\begin{align}
		\phi_{\lambda_1,\lambda_2}^{\sigma \,\, a}(x,\bska)=& \frac{g}{\sqrt{2(2\pi)^3}}\bigg[\frac{x(1-x)}{\kappa_{\perp}^2+m^2(1-x)^2}\bigg]\frac{T^a}{\sqrt{1-x}} \nonumber \\& \chi_{\lambda_1}^{\dagger}\bigg[-\frac{2(\bska\cdot \boldsymbol{\epsilon}_{\lambda_2}^{\perp*})}{1-x}-\frac{1}{x}(\tilde{\sigma}^{\perp}\cdot\bska)(\tilde{\sigma}^{\perp}\cdot \boldsymbol{\epsilon}_{\lambda_2}^{\perp*})+im(\tilde{\sigma}^{\perp}\cdot \boldsymbol{\epsilon}_{\lambda_2}^{\perp*})\frac{1-x}{x}\bigg]\chi_{\sigma} \psi_1^{\sigma},    
	\end{align}
	where, $\phi^{\sigma \,\, a}_{\lambda_1,\lambda_2}(x_i,\bska_i)=\sqrt{P^+}\psi_2 (P,\sigma|k_1,\lambda_1;k_2,\lambda_2)$,  $g$ is the quark-gluon  coupling.
	$T^a$ and $\boldsymbol{ \epsilon}_{\lambda_2}^\perp$ are colour SU(3) matrices and polarization vector of the gluon.
	The quark mass and the two-component spinor for the quark are denoted by $m$ and $\chi_\lambda$ respectively where $\lambda=1,2$ correspond to helicity up/down. 
	We have used the notation $\tilde{\sigma}_1=\sigma_2$ and $\tilde{\sigma}_2=-\sigma_1$ \cite{Harindranath:2001rc, Harindranath:1998ve}.
	Note that, here $x$ and $\bska$ are longitudinal momentum fraction and the relative transverse momentum of quark respectively.
	We define average momentum and the invariant momentum transfer as
	\begin{align}
		P^{\mu}=\frac{1}{2}\left(p^{\prime\mu}+p^{\mu}\right), ~~ \Delta^{\mu}=\left(p^{\prime\mu}-p^{\mu}\right).
	\end{align}
	where, $ p^{\mu} = \bigg(p^+, \boldsymbol{p}^{\perp}, p^{-}\bigg) $ is four momenta in light-front.
	We will calculate the angular momentum distributions in the Drell-Yan frame, defined by $\Delta^+ = 0 \text{ and } \bsP = 0$. Using on-shell conditions $ P \cdot \Delta = 0 \text{ and } P^2 = m^2 - \Delta^2/4$, we get 
	\begin{align}
		\Delta^- = 0, ~~ P^- = \frac{1}{P^+}\left(m^2+\frac{\boldsymbol{\Delta}^{\perp2}}{4}\right).
	\end{align}
	Thus, the initial and the final state four momenta will be
	\be\label{initialmom}
	p^{\mu} = \bigg(P^+, -\frac{\bsD}{2}, \ \frac{1}{P^+}\left(m^2+\frac{\boldsymbol{\Delta}^{\perp2}}{4}\right)\bigg),\\
	\label{finalmom}
	p^{\prime\mu}=\bigg(P^+, \frac{\bsD}{2}, \ \frac{1}{P^+}\left(m^2+\frac{\boldsymbol{\Delta}^{\perp2}}{4}\right)\bigg), 
	\ee
	and the invariant momentum transfer
	\be\label{momtranfer}
	\Delta^\mu\es(p^{\prime}-p)^\mu=\bigg(0, \ \bsD,   0\bigg).
	\ee
	where, $\Delta^2=-\boldsymbol{\Delta}^{\perp2} $.
	
	\section{Angular momentum distributions in front form}

	Now we will calculate the spatial distribution of various components of angular momentum. 
	{{This involves calculating the matrix element of the angular momentum tensor. 
			The matrix element of an orbital angular momentum operator, $x^\nu T^{\mu \rho}-x^\rho T^{\mu \nu}$, when computed between plane wave states presents an issue. 
			This issue arises because the spatial integration of the EMT can result in either an infinite or zero value, leading to ambiguity in its definition \cite{Leader:2013jra}.
			One way to resolve this ambiguity is by considering wave packet states instead of plane wave states.
			However, constructing a proper wave packet for a spin-1/2 particle can be a complex task. 
			One approach is to define the wave packet as a superposition of momentum eigenstates while keeping the spin vector fixed in the rest frame, as suggested in \cite{Bakker:2004ib}. 
			An alternative method to address this problem is to relate the matrix element of the angular momentum operator to the off-forward matrix element of the EMT \cite{Jaffe:1989jz, Shore:1999be}. 
			When this method is correctly applied, it ensures that the spatial integral of the local operator is handled properly, addressing the ambiguity issue.
			Moreover, it preserves the Lorentz covariance of the matrix element of the AM densities and establishes the correct relationship between the matrix element of the EMT and the AM densities \cite{Bakker:2004ib, Leader:2013jra, Lorce:2017wkb}.
			To calculate the spatial distributions of angular momentum correctly, which involves calculating the Fourier transform of the AM densities in the momentum space, we use a Gaussian wave packet state in position space centred at the origin \cite{Diehl:2002he, Chakrabarti:2005zm}. In our previous works, we have used these wave-packet states to calculate the spatial distribution of pressure and shear forces and energy density for a dressed quark state \cite{More:2021stk, More:2023pcy}. The state which is confined in transverse momentum space with definite longitudinal momentum can be written as
			\be
			\frac{1}{16\pi^3}\int \frac{d^2 \bsp dp^+}{p^+}\phi\left( p\right) \mid p^+,\bsp,\lambda \rangle 
			\ee
			with $\phi(p)=p^+\ \delta(p^+-p_0^+)\ \phi\left(\bsp\right)$.
			We choose a Gaussian shape for $\phi\left( \bsp\right)$ in transverse momentum :
			\be\label{gaussian}
			\phi\left(\bsp\right)
			= e^{-\frac{\bspp}{2\sigma^2}}
			\ee 
			where $\sigma$ is the width of Gaussian.
			We use the light-front coordinates to calculate distribution in the transverse plane. This is because the subgroup of Lorentz transformations associated with the transverse plane in light-front coordinates is Galilean in nature \cite{Brodsky:1997de} and hence, no relativistic corrections are required.}}
	
	In LF formalism, the OAM distribution in four-dimensional space is given as \cite{Lorce:2017wkb}
	\begin{align}
		\langle L^z \rangle(x) 
		=& \epsilon^{3jk}x_{\perp}^{j}\int \frac{d\Delta^+d^2\bsD}{(2\pi)^3}e^{i\Delta \cdot x}\langle T^{+k}\rangle_{\text{LF}}, \label{Lz_gen}
	\end{align}
	where $\langle T^{+k}\rangle_{\text{LF}}=\frac{\langle p^{\prime},\boldsymbol{s}|T^{+k}(0)|p,\boldsymbol{s}\rangle}{2\sqrt{p^{\prime+}p^+}}$ is the matrix element of EMT. Now we will use different expressions of $T^{\mu\nu}$ for different decompositions in light-front coordinates and gauge $A^+=0$. In order to calculate the distributions, we considered the in and out states of the matrix elements to be dressed quark states and we only consider the quark part of the EMT in this work. We can re-express Eq. \ref{Lz_gen} in the following way.
	Using the on-shell conditions, we can write
	\begin{align}
		x^j e^{i\Delta.x} &= i\frac{\partial}{\partial \Delta^j} e^{i \Delta.x} + x^+ \frac{P^j}{P^+} e^{i \Delta.x}, \label{convenience}
	\end{align}
	which when substituted in Eq. \eqref{Lz_gen} gives
	\begin{align}
		\left<L^z \right> (\boldsymbol{x}) = \epsilon^{3jk} \int \frac{d\Delta^+d^2\bsD}{(2\pi)^3} \left[ i\frac{\partial}{\partial \Delta^j} e^{i \Delta.x} \right] \left< T^{+k} \right> + \epsilon^{ijk} \int \frac{d\Delta^+d^2\bsD}{(2\pi)^3} x^+ \frac{P^j}{P^+} e^{i\Delta.x} \left< T^{+k} \right>. \nonumber 
	\end{align}
	Using integration by parts for the first term, and ignoring the surface term
	\begin{align}
		\left<L^z \right> (\boldsymbol{x}) = \epsilon^{3jk} \int \frac{d\Delta^+d^2\bsD}{(2\pi)^3}  e^{i \Delta.x} \left[ -i \frac{\partial \left< T^{+k} \right>}{\partial \Delta^j} + x^+ \frac{P^j}{P^+} \left< T^{+k} \right> \right].
	\end{align}
	In the Drell-Yan frame, where the average transverse momentum of the system is zero, the impact-parameter distribution of OAM is given as 
	\begin{align}
		\langle L^z \rangle(\boldsymbol{b}^{\perp})= -i\epsilon^{3jk}\int \frac{d^2\bsD}{(2\pi)^2}e^{-i\bsD \cdot \boldsymbol{b}^{\perp}} \left[\frac{\partial \langle T^{+k}\rangle_{\text{LF}} }{\partial \Delta_{\perp}^{j}}\right], \label{Lz expr}
	\end{align} 
	where $\boldsymbol{b}^{\perp}$ is the impact parameter. It is the Fourier conjugate to the tranverse momentum transfer $\boldsymbol{\Delta_\perp}$. In this frame, the dependence of spatial distribution on light-front time is removed.
	Similarly, the expression for spin distribution in light-front is as follows:
	\begin{align}
		\langle S^z \rangle(\boldsymbol{b}^{\perp})&= \frac{1}{2}\epsilon^{3jk}\int \frac{d^2\bsD}{(2\pi)^2}e^{-i\bsD \cdot \boldsymbol{b}^{\perp}} \langle S^{+jk}\rangle_{\text{LF}}. \label{Sz expr}
	\end{align}
	Also, the Belinfante-improved total angular momentum distribution which has the \text{pure} orbital form is given by
	\begin{align}
		\langle J_{\text{Bel}}^z \rangle(\boldsymbol{b}^{\perp})
		= -i\epsilon^{3jk}\int \frac{d^2\bsD}{(2\pi)^2}e^{-i\bsD \cdot \boldsymbol{b}^{\perp}} \left[\frac{\partial \langle T^{+k}_{\text{Bel}}\rangle_{\text{LF}} }{\partial \Delta_{\perp}^{j}}\right].\label{Jz expr}
	\end{align}
	
	In order to calculate these distributions, we must first evaluate the matrix element of the EMT to which they are associated. First we consider the quark part of the kinetic EMT given by the Eq. \eqref{T_kin} (Ji's decomposition). 
	\begin{align}
		T^{+k}_{\text{kin,q}}(x)=& \frac{1}{2}\overline{\psi}(x)\gamma^{+}i\overleftrightarrow{D}^{k}\psi(x) = \psi_{+}^{\dagger}\left(i\overleftrightarrow{\partial}^{k}+2gA^k\right)\psi_{+}.
	\end{align}
	Using the two-component expressions of fermion and gluon field, we get
	\begin{align}
		\nonumber T^{+k}_{\text{kin,q}}(0)=& \sum_{\lambda,\lambda^{\prime}}\int \frac{dk^{\prime+}d^2k^{\prime \perp}dk^{+}d^2k^{\perp}}{(16\pi^3)^2\sqrt{k^{\prime+}k^+}}b_{\lambda^{\prime}}^{\dagger}(k^{\prime})b_{\lambda}(k) ~ \chi_{\lambda^{\prime}}^{\dagger}\left[k^{\prime k}+k^{k}\right]\chi_{\lambda}\\&+2g \sum_{\lambda^{\prime},\lambda,\lambda_3}\int \frac{dk^{\prime+}d^2k^{\prime \perp}dk^{+}d^2k^{\perp}dk_3^{+}d^2k_3^{\perp}}{(16\pi^3)^3k_3^{+}\sqrt{k^{\prime+}k^+}}~\left[\epsilon_{\lambda_3}^{k}b_{\lambda^{\prime}}^{\dagger}(k^{\prime})b_{\lambda}(k)a_{\lambda_3}(k_3)+\epsilon_{\lambda_3}^{k*}b_{\lambda^{\prime}}^{\dagger}(k^{\prime})b_{\lambda}(k)a^{\dagger}_{\lambda_3}(k_3)\right].
	\end{align}
	We calculate the diagonal (first term) and off-diagonal matrix (second term) elements of $T^{+k}_{\text{kin,q}}(0)$ by sandwiching the above expression between dressed quark state given in Eq. \eqref{state}. These are then expressed in terms of boost-invariant LFWF and Jacobi momenta ($x,\kappa^\perp$).  
	We have shown the form of the diagonal matrix element in Eq. \eqref{T+k_kin1} and \eqref{T+k_kin2} obtained after performing the integration over Jacobi momenta. Substituting these expressions in Eq.\eqref{Lz expr}
	\begin{align}
		\langle L^z_{\text{kin, q}} \rangle (\bsb)=\frac{g^2 C_F}{72 \pi^2} \int \frac{d^2\boldsymbol{\Delta}^{\perp}}{(2\pi)^2}e^{-i\bsb \cdot \bsD} \left[-7+\frac{6}{\omega}\left(1+\frac{2m^2}{\Delta^2}\right)\text{log}\left(\frac{1+\omega}{-1+\omega}\right)-6\text{log}\left(\frac{\Lambda^2}{m^2}\right)\right],
		\label{OAM_kin}
	\end{align}
	where $\omega = \sqrt{1+\frac{4m^2}{\Delta^2}}$ and $\Lambda$ is the ultraviolet cutoff on the transverse momentum. The off-diagonal term, which corresponds to the potential angular momentum vanishes as shown in \ref{appB}. 
	This is a confirmation of an interesting result shown in \cite{Ji:2015sio, Amor-Quiroz:2020qmw}.
	
	For the spatial distribution of the intrinsic part of the quark, we write the operator structure of the Eq.\eqref{S_kin} as
	\begin{align}
		S_{q}^{+jk}=\frac{1}{2} \epsilon^{+jk-}\sum_{\lambda,\lambda^{\prime}}\int \frac{dk^{\prime+}d^2\boldsymbol{k}^{\prime \perp}dk^+d^2\boldsymbol{k}^{\prime \perp}}{\left(16 \pi\right)^2 \sqrt{k^{\prime +}k^+}}b_{\lambda^{\prime}}^{\dagger}(k^{\prime})b_{\lambda}(k) ~ \left(\chi_{\lambda^{\prime}}^{\dagger}\sigma^{(3)}\chi_{\lambda}\right).
	\end{align}
	Again, after evaluating the diagonal and off-diagonal (vanishes) matrix element in terms of Jacobi momenta and substituting them in Eq. \eqref{Sz expr}, we get
	\begin{align}
		\langle S^z_{\text{kin, q}} \rangle(\boldsymbol{b}^{\perp})
		&= -\frac{g^2 C_F}{32 \pi^2}  \int \frac{d^2\bsD}{(2\pi)^2}e^{-i \bsD \cdot \bsb}  \int \frac{dx}{1-x} \times \nonumber \\
		&\hspace{0.4cm}\left[ \omega (1+x^2) \log{\left( \frac{1+\omega}{-1+\omega} \right)} + \left(\frac{1-\omega^2}{\omega}\right) x \log{\left( \frac{1+\omega}{-1+\omega} \right)} - (1+x^2) \log{\left( \frac{\Lambda^2}{m^2(1-x)^2} \right)} \right].
		\label{Spin_kin}
	\end{align}
	
	Similarly, the impact-parameter distribution for the quark part of the Belinfante-improved total angular momentum is calculated using Eq.\eqref{T_Bel} and \eqref{Jz expr}
	\begin{align}
		& \nonumber \langle J_{\text{Bel, q}}^z \rangle(\boldsymbol{b}^{\perp})\\
		&= \nonumber g^2 C_F \int \frac{d^2\boldsymbol{\Delta}^{\perp}}{(2\pi)^2}e^{-i\bsb \cdot \bsD} \int \frac{dx}{16 \pi^2} \frac{1}{\left(1-x\right){\Delta}^4 \omega^3} ~\times\\\nonumber& \bigg[\left(8m^4\left(1-2x\right)\left(1-x\left(1-x\right)\right)+6m^2\left(1-\left(2-x\right)x\left(1+2x\right)\right)\Delta^2+\left(1-\left(2-x\right)x\left(1+2x\right)\right)\Delta^4\right)\log\left(\frac{1+\omega}{-1+\omega}\right)\\&-\omega \Delta^2 \left(4m^2\left(1-\left(1-x\right)x\right)+\left(1+x^2\right)\Delta^2+\left(1-\left(2-x\right)x\left(1+2x\right)\right)\left(4m^2+\Delta^2\right)\log\left(\frac{\Lambda^2}{m^2\left(1-x\right)^2}\right)\right)\bigg].
	\end{align}
	The relevant $\kappa^\perp$ integrations and the steps of the calculation are given in Appen. \ref{appA} and Appen. \ref{appB} respectively.
	
	In order to make a proper comparison between the Belinfante and Ji's decomposition, it is important to include the correction term to the Belinfante's total angular momentum. 
	This term corresponds to the superpotential given in Eq. \eqref{superpotential}. It's expression at the distribution level is given in \cite{Lorce:2017wkb}
	\begin{align}
		\langle M^z \rangle(\boldsymbol{b}^{\perp})&= \frac{1}{2}\epsilon^{3jk}\int \frac{d^2\bsD}{(2\pi)^2}e^{-i\bsD \cdot \boldsymbol{b}^{\perp}} \Delta^l_\perp \frac{\partial \langle S^{l+k}\rangle_{\text{LF}}}{\partial \Delta^j_\perp}.  \label{Mz expr}
	\end{align}
	Substituting the Eq. \eqref{Sl+k} in the above expression, we get the spatial distribution of the superpotential term
	\begin{align}
		\langle M^z_{\text{q}} \rangle(\boldsymbol{b}^{\perp})
		\nonumber=& \frac{g^2C_{F}}{32 \pi^2}\int \frac{d^2\bsD}{(2\pi)^2}e^{-i\bsD \cdot \bsb}\int \frac{dx}{\left(1-x\right)}\frac{1}{\omega^3 \Delta^{4}}\\& \left[\omega \Delta^{2}\left(\left(4m^2+\Delta^{2}\right)\left(1+x^2\right)-4m^2x\right)-2m^2\left(\left(4m^2+\Delta^{2}\right)\left(1+x^2\right)-4m^2x-2x\Delta^{2}\right)\right]. 
	\end{align}
	
	Let us consider now the quark part of canonical EMT given by Eq. \ref{T_can}. For the spin distribution in the canonical definition, we use the 2nd term of Eq. \eqref{S_can}. Since the non-diagonal matrix element of kinetic EMT which is due to the presence of the gauge field in the covariant derivative, is found to be zero (see Appen. \ref{appB}), so effectively, we get the non-zero contribution only from
	\begin{align}
		T^{+k}_{\text{kin, q}}= \frac{1}{2}\overline{\psi}\gamma^+ i\overleftrightarrow{\partial}^{k}\psi = \psi_{+}i\overleftrightarrow{\partial}^{k}\psi_{+} = T^{+k}_{\text{q}},
	\end{align}
	which is exactly equal to the canonical definition of EMT in light-front. Thus the OAM distribution of kinetic and canonical is the same. Also, the spin density is the same as the kinetic one, see Eq. \ref{S_can} and \ref{S_kin}. Thus the orbital angular momentum distribution and the spin distribution for the quark in canonical decomposition is 
	\begin{align}
		\langle L^z_{\text{kin, q}} \rangle(\boldsymbol{b}^{\perp}) = \langle L^z_{\text{q}} \rangle(\boldsymbol{b}^{\perp}), \hspace{3cm} \langle S^z_{\text{kin, q}} \rangle(\boldsymbol{b}^{\perp}) = \langle S^z_{\text{q}} \rangle(\boldsymbol{b}^{\perp})
	\end{align}
	
	Considering the quark part of gic decomposition of EMT, see Eq. \ref{T_gic}, which gives Chen et al. decomposition \cite{Chen:2008ag, Chen:2009mr,Leader:2013jra}, we get
	\begin{align}
		T^{+k}_{\text{gic, q}}= \frac{1}{2} \overline{\psi} (x) \gamma^{+} i\overleftrightarrow{D}^{k}_{\text{pure}} \psi (x),
	\end{align}
	where, $\overleftrightarrow{D}^{\mu}_{\text{\text{pure}}}=\overleftrightarrow{\partial}^{\mu}-2igA^{\mu}_{\text{\text{pure}}}$ and $A_{\text{\text{pure}}}^{\mu}=A^{\mu}-A_{\text{\text{phys}}}^{\mu}$. Since only the transverse component of the gauge field in the light-front is dynamical, i.e., $A^{k}=A_{\text{\text{phys}}}^{k}$ and in the light-front gauge $A^+ = 0$. Thus, the gic decomposition of EMT is the same as the canonical one,
	\begin{align}
		T^{+k}_{\text{gic, q}}= \frac{1}{2} \overline{\psi} (x) \gamma^{+} i\overleftrightarrow{\partial}^{k} \psi (x)=T^{+k}_{\text{q}}.
	\end{align}
	Thus OAM distribution calculated from the gic decomposition of EMT is the same as given in Eq. \ref{OAM_kin}. The same goes for the spin distributions can be seen from the second term of Eq. \ref{J_gic}. So spin distribution due to this is given by Eq. \ref{Spin_kin} and \ref{S_kin}.
	\begin{align}
		\langle L^z_{\text{gic, q}} \rangle(\boldsymbol{b}^{\perp}) = \langle L^z_{\text{q}} \rangle(\boldsymbol{b}^{\perp}), \hspace{3cm} \langle S^z_{\text{gic, q}} \rangle(\boldsymbol{b}^{\perp}) = \langle S^z_{\text{q}} \rangle(\boldsymbol{b}^{\perp}).
	\end{align}
	
	We can also see from  Eq. \ref{T_gik} that the quark part of Wakamatsu's decomposition \cite{Wakamatsu:2010cb, Wakamatsu:2010qj} is similar to the quark part of Ji's decomposition \ref{T_kin}. So from this decomposition, we will not get any new information
	\begin{align}
		T^{\mu\nu}_{\text{gik, q}}=T^{\mu\nu}_{\text{kin, q}}.
	\end{align}
	The intrinsic part is also exactly the same as the kinetic one, see Eq. \ref{S_kin} and the second term of \ref{J_gik}. So, at the distribution level, the gik decomposition is given as 
	\begin{align}
		\langle L^z_{\text{gik, q}} \rangle(\boldsymbol{b}^{\perp}) = \langle L^z_{\text{kin, q}} \rangle(\boldsymbol{b}^{\perp}), \hspace{3cm} \langle S^z_{\text{gik, q}} \rangle(\boldsymbol{b}^{\perp}) = \langle S^z_{\text{kin, q}} \rangle(\boldsymbol{b}^{\perp}).
	\end{align}
	
	\section{Extraction of gravitational form factor D}
	If we consider a general asymmetric EMT, then its matrix elements for a spin$-\frac{1}{2}$ target are parametrized by five form factors \cite{Bakker:2004ib}
	\begin{align}
		\nonumber &\langle p^{\prime}, \bs{s}^{\prime}|T^{\mu\nu}(0)| p, \bs{s} \rangle \\\nonumber =& \overline{u}\left(p^{\prime}, \bs{s}^{\prime}\right)\left[\frac{P^{\mu}P^{\nu}}{M}A(t)+\frac{P^{\mu}i\sigma^{\nu\lambda}\Delta_{\lambda}}{4M}\left(A+B+D\right)(t)+\frac{\Delta^{\mu}\Delta^{\nu}-g^{\nu\nu}\Delta^{2}}{M}C(t)+mg^{\mu\nu}\overline{C}(t)\right. \\ & \left.+\frac{P^{\nu}i\sigma^{\mu\lambda}\Delta_{\lambda}}{4M}\left(A+B-D\right)(t) \right]u\left(p,\bs{s}\right), 
	\end{align}
	where $M$ is the nucleon mass, $\bs{s}, \bs{s}^{\prime}$ denotes the rest frame polarization of the initial and the final states respectively, and
	\begin{align}
		P=\frac{p^{\prime}+p}{2}, ~~~ \Delta= p^{\prime}-p, ~~~ t=\Delta^2.
	\end{align}
	The matrix element of the quark spin density operator is also parameterized by two form factors
	\begin{align}
		\langle p^{\prime}, \bs{s}^{\prime}|S^{\mu\alpha\beta}(0)| p, \bs{s} \rangle = \frac{1}{2}\epsilon^{\mu\alpha\beta\lambda}~ \overline{u}(p^{\prime},\bs{s}^{\prime})\left[\gamma_{\lambda}\gamma_{5}G^{q}_{A}(t)+\frac{\Delta_{\lambda}\gamma_{5}}{2M}G^{q}_{P}(t)\right]u(p,\bs{s}), 
	\end{align}
	where $G^{q}_{A}(t)$ and $G^{q}_{P}(t)$ are axial vector and pseudoscalar form factors respectively. Usually, axial vector form factors are isoscalar (u-d like state) or isovector (u-u like state) in nature but since we are considering a state with only one quark dressed with a gluon, we will address $G^{q}_{A}(t)$ simply as axial vector form factor.
	The axial vector form factor is measurable
	from quasielastic neutrino scattering and pion {electro-production processes} \cite{Bernard:2001rs}.
	The antisymmetric part of the EMT is related to the divergence of the spin density $T^{\mu\nu}-T^{\nu\mu}=-\partial_{\mu}S^{\mu\alpha\beta}(x)$ and the antisymmetric part of the quark EMT is given by
	\begin{align}
		\overline{\psi}(x)\left[\gamma^{\alpha}i\overleftrightarrow{D}^{\beta}-\gamma^{\beta}i\overleftrightarrow{D}^{\alpha}\right]\psi(x)=-\epsilon^{\mu\alpha\beta\lambda}\partial_{\mu}\left[\overline{\psi}(x)\gamma_{\lambda}\gamma_{5}\psi(x)\right].
	\end{align}
	So the form factor associated with the antisymmetric part of the quark EMT is the axial vector form factor
	\begin{align}
		D_{q}(t)= -G^{q}_{A}(t).
	\end{align}
	
	To extract this form factor in the dressed quark state, we use the following relation 
	\begin{align}
		\frac{\langle p^{\prime}, \bs{s}|S_{\text{q}}^{+jk}(0)| p, \bs{s} \rangle}{2p^{+}}=\frac{1}{2}\epsilon^{+jk-} s^{z}~ G^{q}_{A}(t),
	\end{align}
	where we have used the relation $\overline{u}(p^{\prime},\bs{s}^{\prime})\gamma^{+}\gamma_{5}u(p,\bs{s})=$4$P^{+}s^{z}$ in Drell-Yan frame with light-front spinors by taking $n_{\text{LF}}=\left(1,0,0,-1\right)$\cite{Lorce:2017isp}.
	After comparing the left-hand side of the above equation with Eq. \ref{Sl+k}, we get 
	\begin{align}
		D_{q}(t)=& -  g^2C_F \int \frac{dxd^2\bska}{16\pi^3} \frac{1}{(1-x)D_1D_2}\times \nonumber \\
		& \hspace{0.4cm} \left[ {\kappa^{\perp2}} (1 + x^2) + \bska \cdot \bsD (1 - x)(1 + x^2) + i(1-x)(1 - x^2) (\kappa^{(1)}\Delta^{(2)}-\kappa^{(2)}\Delta^{(1)}) - m^2(1 - x)^4\right],
	\end{align}
	where  \begin{align}
		D_1 =& \bskasq+m^2\left(1-x\right)^2, \\
		D_2=& \left(\bska+\left(1-x\right)\bsD\right)^2+m^2\left(1-x\right)^2.
	\end{align}
	After performing the $\kappa$ integration, we get:
	\begin{align}
		D_{q}(t) = &-\frac{g^2 C_F}{16 \pi^2}  \int \frac{dx}{1-x} \left[ \omega (1+x^2) \log{\left( \frac{1+\omega}{-1+\omega} \right)} + \left(\frac{1-\omega^2}{\omega}\right) x \log{\left( \frac{1+\omega}{-1+\omega} \right)} - (1+x^2) \log{\left( \frac{\Lambda^2}{m^2(1-x)^2} \right)} \right],
	\end{align}
	where $\omega = \sqrt{1+\frac{4m^2}{\Delta^2}}$. 
	
	\section{Numerical analysis}
	
	In this section, we show the {{numerical results of the longitudinal component of the angular momentum distribution 
			in the dressed quark state}}. The {{analytic results}} in the previous section {{show}} that the AM distribution for quarks is the same for Ji, JM, gic, and gik decompositions. So, we only plot the results for Belinfante and Ji decomposition.
	For the analysis, we have chosen the parameters: the quark mass $m = 0.3$ GeV, the coupling constant $g = 1$, the colour factor $C_F = 1$, and the ultraviolet cutoff $\Lambda = 1.7$ GeV. 
	Also, the y-axis is multiplied by a factor of ${b}_\perp$ to correctly represent the data in radial coordinates where ${b}_\perp$ is the modulus of vector $\bs{b}_{\perp}$. {{In order to perform the Fourier transform of the different components of angular momentum density and obtain smooth plots for spatial distribution in the impact-parameter space, we have chosen Gaussian wave packets of width $\sigma$ instead of plane waves.}}
	
	\begin{figure}[ht]
		\begin{minipage}{0.49\linewidth}
			\includegraphics[scale=0.95]{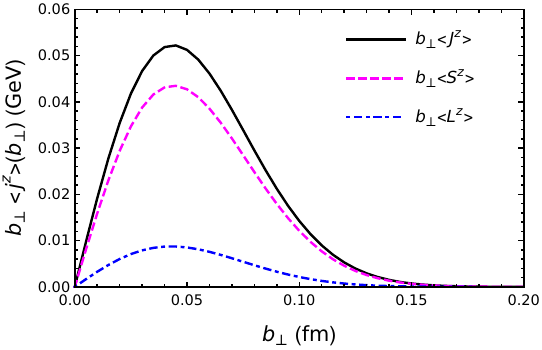}
		\end{minipage}
		\begin{minipage}{0.49\linewidth}
			\includegraphics[scale=0.95]{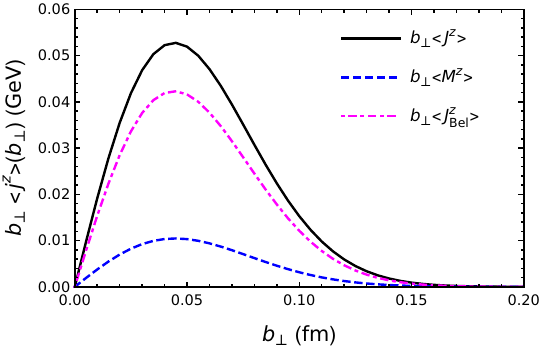}
		\end{minipage}  
		\caption{(Colour) Plots of longitudinal angular momentum distribution of quarks as a function of impact parameter ${b}_\perp$. Left: Sum of the kinetic orbital AM ${b}_\perp\langle L^z \rangle$ (dot-dashed line) and spin AM ${b}_\perp \langle S^z \rangle$ (dashed line) given by kinetic total AM ${b}_\perp \langle J^z \rangle$ (solid line). Right: Kinetic total AM ${b}_\perp \langle J^z \rangle$ (solid line) is given by the sum of Belinfante total AM ${b}_\perp \langle J^z_{\text{Bel}} \rangle$ (dot-dashed line) and the correction term corresponding to the total divergence ${b}_\perp \langle M^z \rangle$ (dashed line). Here, $m = 0.3$ GeV, $g=1$, $C_f=1$, and $\Lambda = 1.7$ GeV. We choose the Gaussian width $\sigma =0.1 $ GeV.}
		\label{Dist_graph}
	\end{figure}
	
	\begin{figure}[ht]
		\centering
		
		\begin{minipage}{0.45\textwidth}
			\centering
			\includegraphics[scale=0.87]{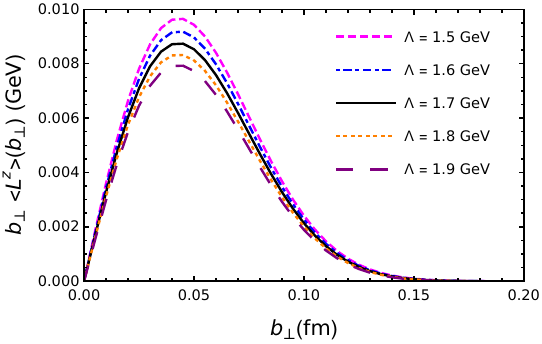}
		\end{minipage}
		\hfill
		\begin{minipage}{0.45\textwidth}
			\centering
			\includegraphics[scale=0.87]{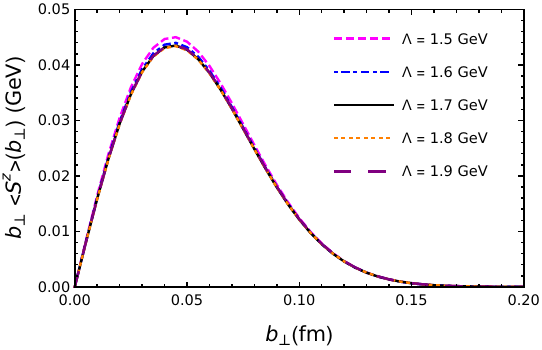}
		\end{minipage}
		
		\medskip

		\centering
		\includegraphics[scale=0.87]{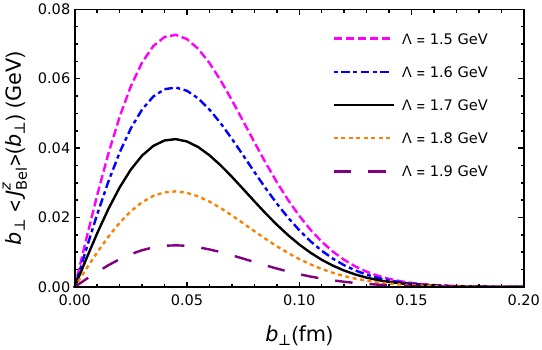}
		
		\hfill
		
		\caption{(Colour) Plot of the dependence of different components of AM distribution on transverse momentum UV cutoff. Top-left: Variation of $b_\perp \langle L^z_{\text{kin, q}} \rangle$ with $\Lambda$. Top-right: Variation of $b_\perp \langle S^z_{\text{kin, q}}\rangle$ with $\Lambda$. Bottom: Variation of $b_\perp \langle J^z_{\text{Bel, q}} \rangle$ with $\Lambda$. Five different values of $\Lambda$ are considered for the analysis: $\Lambda = 1.5,1.6,1.7,1.8,1.9$ GeV. The Gaussian width is $\sigma = 0.1$ GeV.}
		\label{Cutoff_graph}
	\end{figure}
	
	In Fig. \ref{Dist_graph}, the plot on the left panel {{shows}}  the longitudinal component of Eq. \eqref{Ji_graph} in the impact-parameter space, $\langle J^z_{\text{kin,q}} \rangle (b_\perp) = \langle L^z_{\text{kin,q}} \rangle (b_\perp) + \langle S^z_{\text{kin,q}} \rangle (b_\perp)$. Both the OAM and spin distribution have a positive contribution. 
	{{Unlike the analysis done in a scalar diquark model in  \cite{Lorce:2017wkb},}} {{without a gluon contribution}}, with our dressed quark state, we find that the spin contribution to the total angular momentum of quarks is dominating over the OAM contribution. 
	This result is similar to the analysis done for a proton in \cite{Liu:2022fvl} using the basis light-front quantization approach, {{where the valence sector LFWF of the proton have been considered.}}
	{{A similar study of AM distribution done by using light-front quark-diquark LFWFs in \cite{Kumar:2017dbf} also shows that the intrinsic contribution to the total angular momentum is {{dominant}} in comparison to the OAM contribution.}}
	{{It would be interesting to see the effect of higher Fock components in the contribution of quark OAM. From the right panel of Fig. \ref{Dist_graph}, it can be seen}} that the Belinfante total AM distribution, $\langle J^z_{\text{Bel,q}} \rangle (b_\perp)$ is not equal to the total AM density of the quark.  A correction term corresponding to the superpotential, $\langle M^z_{\text{q}} \rangle (b_\perp)$, which is ignored in a symmetric EMT like the Belinfante, has to be added to the  
	$\langle J^z_{\text{Bel,q}} \rangle (b_\perp)$. 
	However, in our analysis, the contribution of this superpotential is positive for the whole range of impact parameters. This contrasts with the results of \cite{Liu:2022fvl, Lorce:2017wkb} where the correction term has a positive contribution near the core but a negative contribution near the periphery. {{However, the major contribution comes from the $\langle J^z_{\text{Bel,q}} \rangle (b_\perp)$ and only a small fraction of the $\langle J^z \rangle (b_\perp)$ is attributed to $\langle M^z \rangle (b_\perp)$ which is in agreement with the analysis of Belinfante decomposition in \cite{Lorce:2017wkb}.}}
	
	{{As discussed above, the angular momentum derived from the canonical, Belinfante and kinetic AM tensors will be the same, as they differ by a surface term which goes away upon integration, with the assumption that the fields vanish at infinity. However, individually the contributions coming from the quark and gluon part of the AM tensor are not conserved, this is not expected unless the full AM tensor is considered. In this work, we have considered only the quark part of the AM tensor and the interaction term. When a non-trivial interaction is present in the system, like it is in the dressed quark state, quark and gluon parts of the angular momentum $J_i$ depend on the renormalization scale \cite{Ji:1995cu, Ji:2010zza, Altarelli:1977zs}. In our approach, the scale dependence comes from the cutoff on the transverse momentum integration}}\cite{More:2021stk, More:2023pcy}. 
	In Fig. \ref{Cutoff_graph}, we show the cutoff dependence of different components of the AM density for the dressed quark state.  We see that Belinfante's total AM density shows the largest variations with respect to the transverse momentum cutoff.
	{{Also, as seen in the analytic expression, the superpotential term is independent of the cutoff.}} {{The peaks of all the distributions shift to lower values with an increase in the value of the cutoff. }}
	However, the shape of the distributions doesn't change and all of them have their peaks at around $b_\perp \approx 0.045$ fm. {{Thus, for the quark part of AM tensor, in general, the equality $J_{\text{kin, q}}^{z}=J^{z}_{\text{Bel, q}}+M^{z}$ is not expected to hold, and each side is cutoff dependent.  The contribution coming from the gluon part of the AM tensor needs to be incorporated to verify this equality, which is an ongoing work. In this work, we have chosen a cut-off $\Lambda=1.7$ GeV for which this equality holds.}}
	{{{Fig. \ref{Gaussian-width} shows the variation of the results with the width of the wave packet state, $\sigma$. The width indicates the spread of the distribution.  We observe that for all the components of angular momentum in both Belinfante and Ji decomposition, the peaks of the distributions shift away from the center, and the distributions become broader in $b_\perp$  space with an increase of the width of the Gaussian. }}
		
		In Fig. \ref{Dformfactor}, we plot the {{axial vector form-factor $G^q_A(\Delta^2_\perp)/G^q_A(0)$  as a function of squared momentum transfer $\Delta^2_\perp$. 
				We observe that this form factor is positive over the whole range of $\Delta^2_\perp$  having the maximum at $\Delta^2_\perp=0$.}}
		Our calculations show that $D_q (0) = -0.36$ for cutoff $\Lambda = 1.7 $ GeV. 
		This form factor is directly related to the axial-vector form factor $G_A^q$ i.e., $D_q(\Delta^2_\perp) = - G_A^q(\Delta^2_\perp)$ \cite{Lorce:2018egm}.  
		Thus, the value of the axial-vector form factor of the quark in the dressed quark state is $G_A^q(0) = 0.36$. {{For t=0, leading-twist NNLO analysis performed by HERMES collaboration shows that $ G_A^q(0) = 0.33 \pm 0.011 (\text{theo}) \pm 0.025 (\text{exp}) \pm 0.028 (\text{evol})$ \cite{PhysRevD.75.012007}}}. {{The form factor $G_A$ is also calculated in \cite{Chen:2021guo, Chen:2020wuq} for a quark + scalar diquark (spectator) system and it is found that $G_A(0) = 0.71 \pm 0.04$. However, after considering contributions from other possible diagrams involving axial vector diquarks, $G_A (0)$ is found to be close to 1.25.}} {{Qualitatively, the behavior of $G^q_A(\Delta^2_\perp)/G^q_A(0)$ as a function of $\Delta_\perp^2$ is similar to many other model-based analysis done in \cite{Silva:2005fa, Chen:2021guo, Chen:2020wuq, Dahiya:2017fmp, Ma:2002xu} and lattice studies \cite{Jang:2019vkm, RQCD:2019jai}. The difference in the numerical values can be linked to the fact that these models and lattice studies consider a nucleon state whereas we consider a dressed quark state.}} In the impact-parameter space, $D_q(b_\perp)$ is related to the first Mellin moment of the helicity-dependent generalized parton distribution $\tilde{\mathcal{H}}(x,0,b_\perp)$ \cite{diehl2003generalized} which is given by $\int dx \tilde{\mathcal{H}}(x,0,b_\perp) = - D_q(b_\perp) = G_A^q (b_\perp)$.
		
		{{The variation of the z-component of AM distributions with the two components of the impact-parameter space, $b_x$ and $b_y$, is represented in a 3D plot in Fig. \ref{3Dplots}. We have plotted the magnitude of the Fourier transform in impact-parameter space.
				We observe that the maximum contribution to the total angular momentum density in Ji's decomposition comes from the intrinsic part. We also observe that $\langle S^z_{\text{kin,q}} \rangle$ falls faster than $\langle L^z_{\text{kin,q}} \rangle $.
				The qualitative behavior of the AM distribution plots of $\langle J^z_{\text{Bel,q}} \rangle$ and $\langle M^z \rangle $ in Fig. \ref{3Dplots} is similar to that of $\langle S^z_{\text{kin,q}} \rangle$ and $\langle L^z_{\text{kin,q}} \rangle $ respectively.}}
		
		\begin{figure}[ht]
			\centering
			\begin{minipage}{0.45\textwidth}
				\centering
				\includegraphics[scale=0.87]{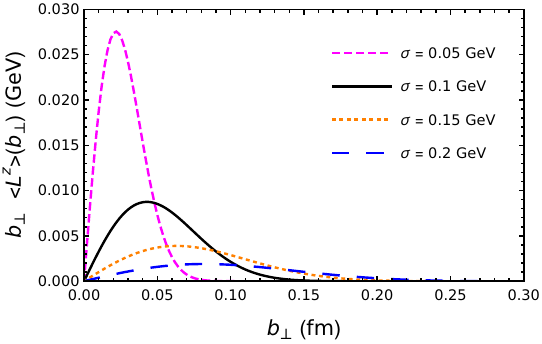}
			\end{minipage}
			\hfill
			\begin{minipage}{0.45\textwidth}
				\centering
				\includegraphics[scale=0.87]{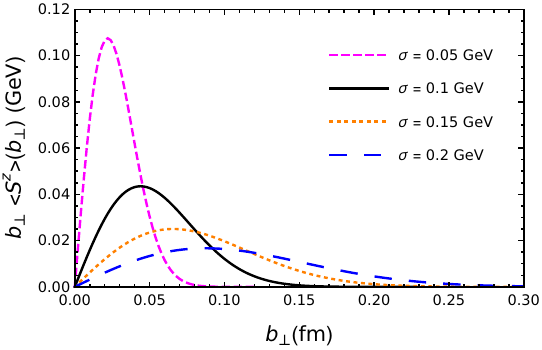}
			\end{minipage}
			
			\medskip
			
			\begin{minipage}{0.45\textwidth}
				\centering
				\includegraphics[scale=0.87]{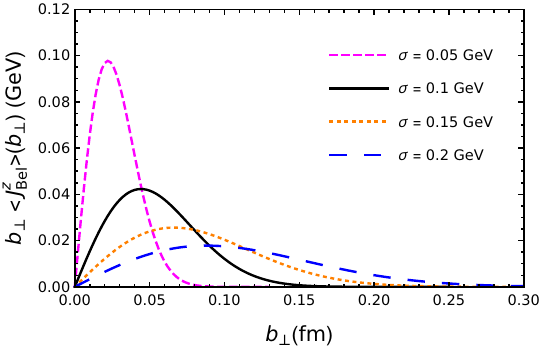}
				
			\end{minipage}
			\hfill
			\begin{minipage}{0.45\textwidth}
				\centering
				\includegraphics[scale=0.87]{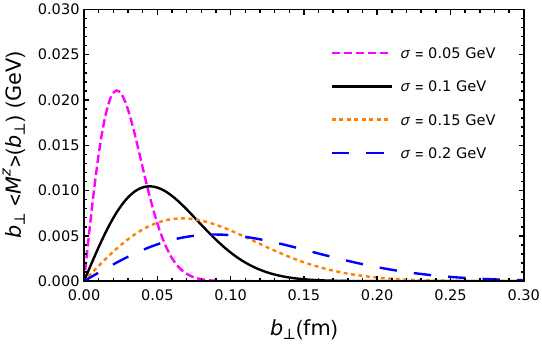}
			\end{minipage}
			
			\caption{(Colour) Plots showing the dependence of different components of AM distribution on the width of the Gaussian wave-packet, $\sigma$. Top-left: Variation of $b_\perp \langle L^z_{\text{kin, q}} \rangle$ with $\sigma$. Top-right: Variation of $b_\perp \langle S^z_{\text{kin, q}}\rangle$ with $\sigma$. Bottom-left: Variation of $b_\perp \langle J^z_{\text{Bel, q}} \rangle$ with $\sigma$. Bottom-right: Variation of $b_\perp \langle M^z_{\text{q}} \rangle$ with $\sigma$. Four different values of $\sigma$ are considered for the analysis: $\sigma = 0.05,0.10,0.15,0.20$ GeV. Here, $m=0.3$ GeV, $\Lambda=1.7$ GeV, and $g=C_f=N_f=1$.}
			\label{Gaussian-width}
		\end{figure}
		
		\begin{figure}[ht]
			\centering
			\includegraphics[scale=0.8]{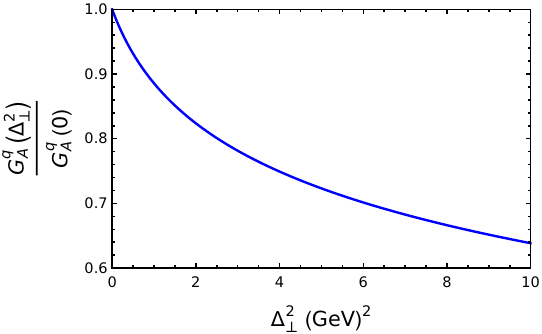}
			\caption{(Colour) Plot of the axial vector form-factor $\frac{G^q_A(\Delta^2_\perp)}{G^q_A(0)}$ as a function of $\Delta^2_\perp$. Here $m = 0.3$ GeV, $g = 1$, $\Lambda = 1.7 $ GeV and  $C_f = 1 $. }
			\label{Dformfactor}
		\end{figure}
		
		\begin{figure}[ht]
			\centering
			
			\begin{minipage}{0.45\textwidth}
				\centering
				\includegraphics[scale=0.47]{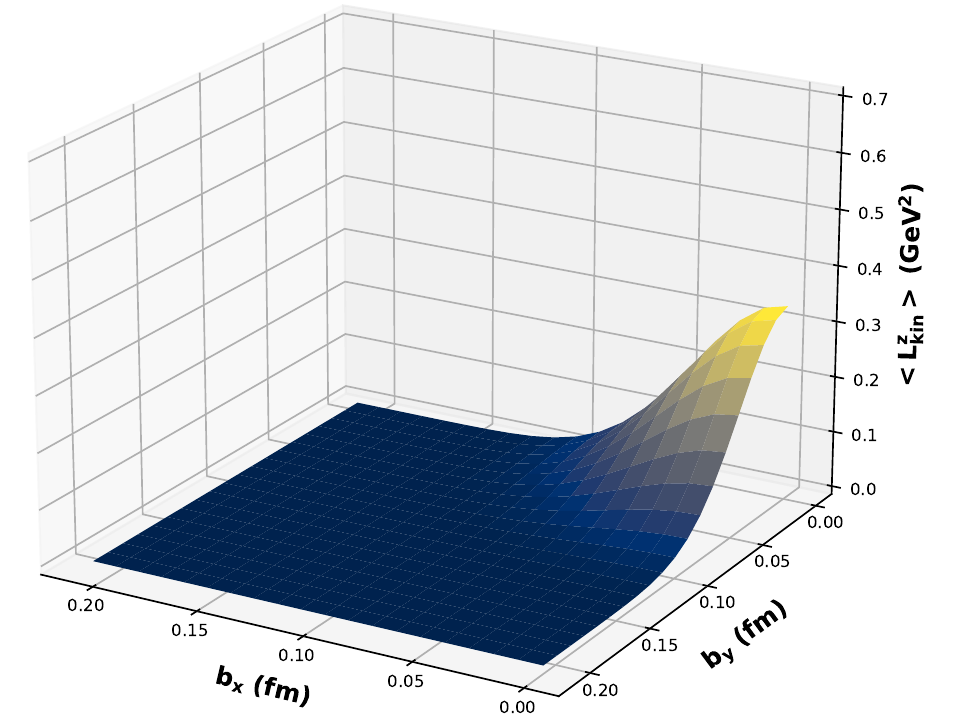}
			\end{minipage}
			\hfill
			\begin{minipage}{0.45\textwidth}
				\centering
				\includegraphics[scale=0.47]{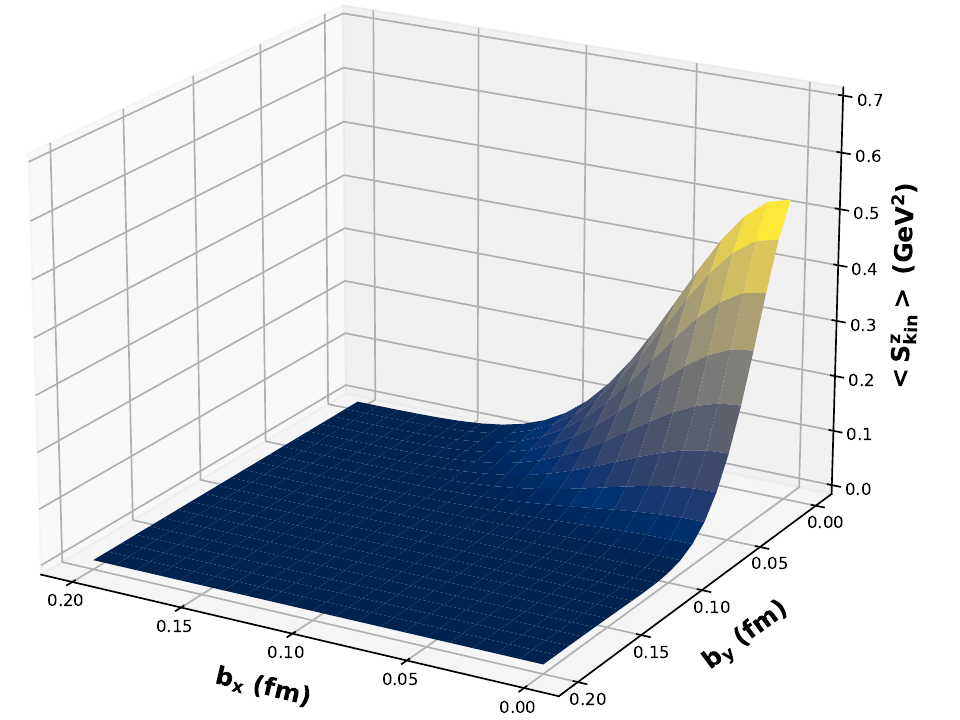}
			\end{minipage}
			
			\medskip
			
			\begin{minipage}{0.45\textwidth}
				\centering
				\includegraphics[scale=0.47]{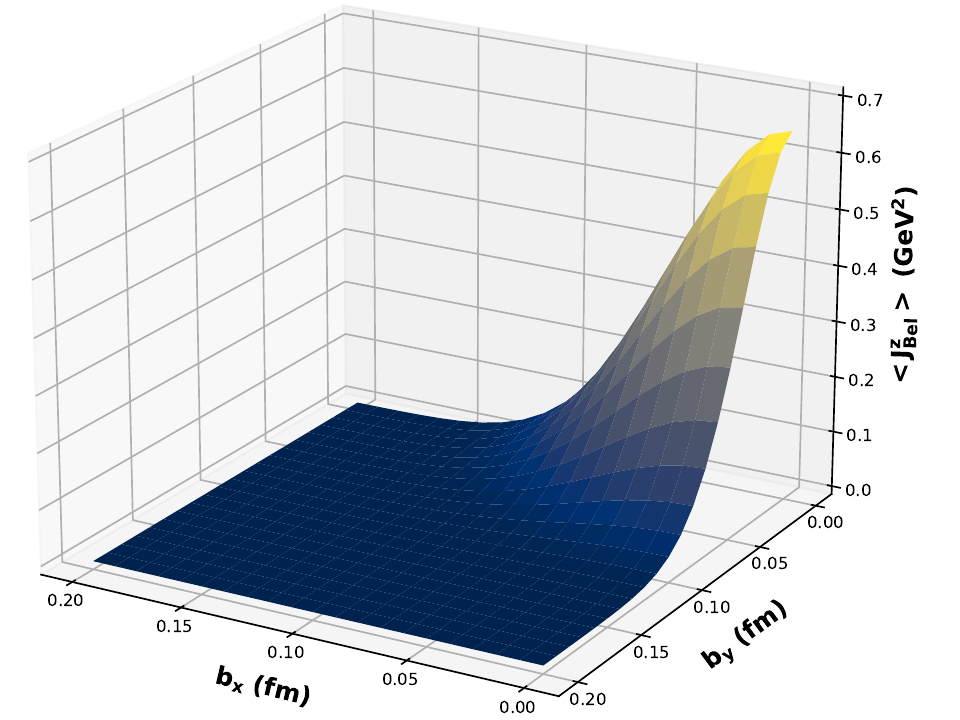}
			\end{minipage}
			\hfill
			\begin{minipage}{0.45\textwidth}
				\centering
				\includegraphics[scale=0.47]{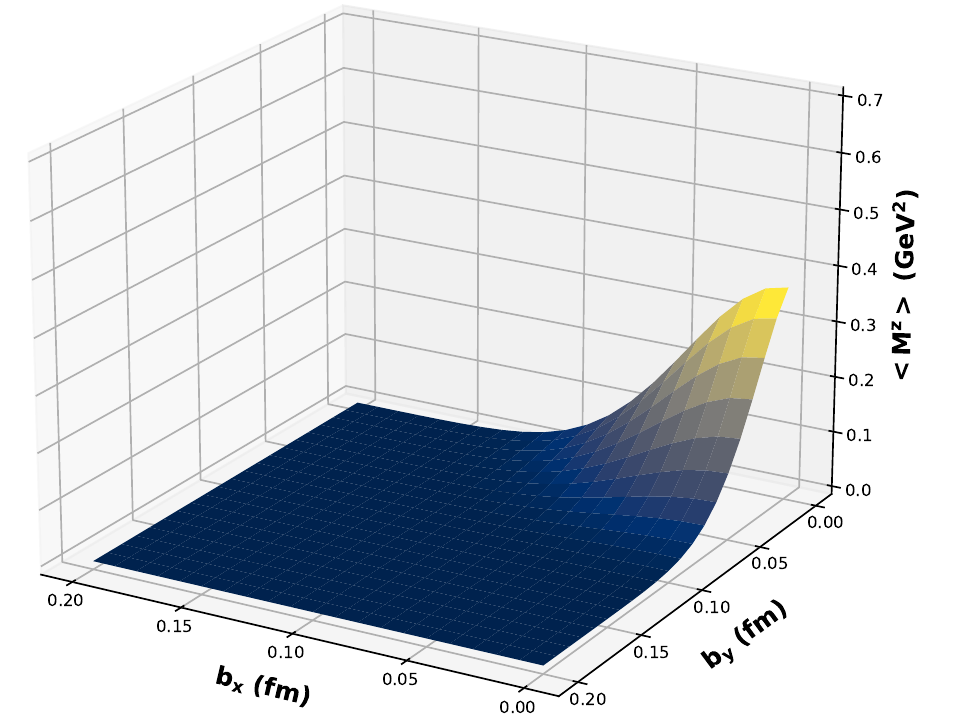}
			\end{minipage}
			
			\caption{(Colour) 3D plot of the expectation value of angular momentum distributions as a function of impact parameters $b_x \text{ and } b_y$. Top-left: Spatial distribution of quark OAM in Ji and Jaffe-Manohar decomposition. Top-right: Spatial distribution of quark spin in Ji and Jaffe-Manohar decomposition. Bottom-left: Spatial distribution of quark OAM in Belinfante decomposition. Bottom-right: Spatial distribution of the superpotential term. We use the following parameters: $m=0.3$ GeV, $g=C_f=N_f=1$, $\Lambda=1.7$ GeV, and the Gaussian width, $\sigma=0.1$ GeV.}
			\label{3Dplots}
		\end{figure}
		\begin{figure}[ht]
			\centering
			\includegraphics[scale=0.47]{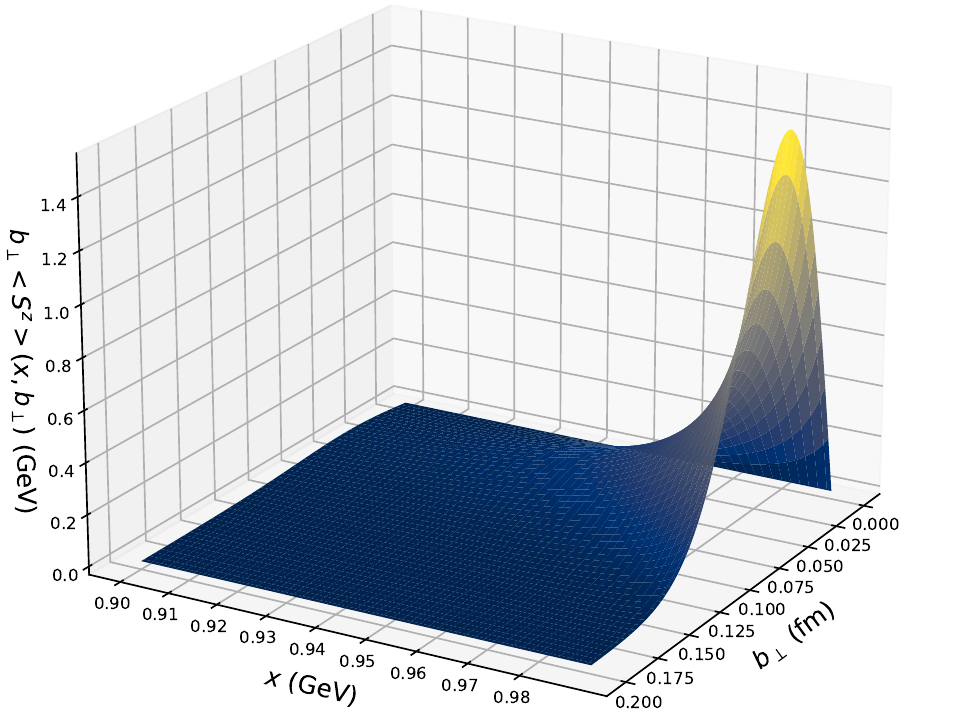}
			\caption{(Colour) Plot of variation of $b_\perp \langle S^z \rangle (x, b_{\perp})$  as a function of longitudinal momentum fraction, $x$ and impact parameter, $b_\perp$. Here $m = 0.3$ GeV, $g=C_f=N_f=1, \Lambda = 1.7$ GeV, and $\sigma=0.1$ GeV.} 
			\label{GPDplot}
		\end{figure}
		We {{show}} the longitudinal component of quark spin AM density, $\langle S^z \rangle$ a function of the longitudinal momentum fraction, $x$, and the impact parameter, $b_\perp$ in Fig. \ref{GPDplot}.
		{{This is called impact parameter dependent parton distribution}} and is given by  Fourier transform of the helicity-dependent GPD,  $\tilde{\mathcal{H}}(x,0,\Delta_\perp)$. For the polarization considered here, 
		\begin{align}
			2\int dx \,\, S^z (x, b_\perp) = G_A (b_\perp) = \int dx \,\, \tilde{H} (x,0,b_\perp)
		\end{align}
		We plot $b_\perp \langle S^z \rangle$ on the y-axis instead of $\langle S^z \rangle$ since $b_\perp$ is the Jacobian associated with transformation of $\langle S^z \rangle$ to polar coordinates. {{Though the GPDs don't have a probabilistic 
				interpretation, but for zero skewness the impact parameter dependent parton distributions have a probabilistic interpretation \cite{Burkardt:2000za}.}} From Fig. \ref{GPDplot}, it is clear that {{contribution to the $\langle S^z \rangle$ only comes from the region of very large momentum fraction for the whole range of the $b_\perp$ considered here. The width of the distribution decreases with a decrease in the value of x. This behaviour of the GPD $\tilde{H} (x,0,b_\perp)$ is different from that shown in \cite{Liu:2022fvl, Freese:2020mcx}}}. {{This difference can be attributed to the fact that in \cite{Liu:2022fvl} the light-front wave functions of a  nucleon state is considered in the valence sector, and in \cite{Freese:2020mcx} Nambu–Jona-Lasinio  (NJL) model is used for the nucleon;   whereas in this work we consider a dressed quark state at one loop.}}
		
		\section{Conclusions}
		
		In this work, we have calculated the spatial distribution of angular momentum for various decompositions and the GFF associated with the antisymmetric part of the EMT. {{At the density level one can investigate the effect of total divergence terms that vanish after integration. Also, gluons contribute significantly to the total angular momentum. Most of the studies in the literature related to the spatial distribution of angular momentum components have been done using phenomenological models of nucleon states that do not include gluons.}} In this work, instead of a nucleon state, we consider a relativistic spin-1/2 state comprising a quark dressed with a gluon, treated perturbatively at the one-loop level in QCD. This state, which incorporates a gluonic degree of freedom, facilitates analytical calculations of quark-gluon light-front wave functions (LFWFs) using the light-front Hamiltonian QCD. The LFWFs are frame-independent due to their expression in terms of relative momenta. {{Employing a two-component formalism in the light-front gauge, we explore the angular momentum and spin distributions as well as the gravitational form factor associated with the antisymmetric part of the EMT}}, all derived from overlaps of these two-particle LFWFs. {The advantage of the light-front gauge} lies in eliminating constrained degrees of freedom, {{allowing for a calculation of the different components of EMT  relevant for the angular momentum distribution}}. Consequently, using the LFWFs of the dressed quark state, we have done a comprehensive analysis of the spatial density of the longitudinal component of intrinsic and orbital angular momentum in the light-front gauge using different definitions of AM decomposition.  
		We have observed that the spatial distribution of total angular momentum when calculated using Belinfante and Ji decomposition does not agree.
		We have calculated the missing superpotential term in the dressed quark state at the distribution level which is responsible for this disparity. {{In the work, we presented the contribution coming from the quark part of the EMT; the contribution from the gluon part to these observables will be discussed in a separate publication. }} Our investigation sheds light on the difference between the Ji and Jaffe-Manohar definitions of angular momentum.
		We found that the potential angular momentum, responsible for this distinction, vanishes at the one-loop level in the dressed quark state. 
		This observation supports previous findings, emphasizing that potential angular momentum arises from the torque exerted on the quark due to the Lorentz force.
		As such, it becomes unmeasurable in a two-body system where the constituents and their center of inertia are all connected by a single line. 
		Consequently, Ji and Jaffe-Manohar's decomposition yields consistent results.
		Since we are working in a light-front gauge where the physical part of the gauge field coincides with the full gauge potential, even Wakamatsu and Chen et al. decomposition gives the same result as Ji and JM  decomposition. 
		We also {{calculated}} the D form factor for quark which is directly related to the axial vector form factor and is associated with the antisymmetric part of the energy-momentum tensor. {{We have compared our results with other calculations in the literature}}.
		\section{Acknowledgement}
		We acknowledge the funding from Board of Research in Nuclear Sciences (BRNS), Government of India, under sanction 
		No. 57/14/04/2021-BRNS/57082. A. M. would like to thank SERB-POWER Fellowship (SPF/2021/000102) for financial support.
		
		\appendix
		
		\section{Integrals used to calculate GFFs}\label{appA}	
		
		The following integrals are used to calculate the analytical forms of the GFFs
		
		\begin{align*}
			&\int d^2\bska \frac{1}{D_1} = \pi \log\left[\frac{\Lambda^2+ m^2(1-x)^2}{m^2(1-x)^2}\right] , \tag{A1}\\
			&\int d^2\bska \frac{1}{D_1 D_2} = \frac{\pi}{(1-x)^2} \frac1{q^2}\ \frac{f_2}{ f_1} , \tag{A2}\\
			&\int d^2\bska \frac{\kappa^{(i)}}{D_1 D_2} = -\frac{\pi}{(1-x)}\frac{ q^{(i)}}{q^2 }\frac{f_2}{2 f_1}, \tag{A3}\\
			&\int d^2\bska \frac{\kappa^{(1)}\kappa^{(2)}}{D_1 D_2}= \pi\frac{q^{(1)}q^{(2)}}{q^2}\bigg[-1+\left(1+\frac{2m^2}{q^2}\right)\frac{f_3}{2f_1}\bigg], \tag{A4}\\
			&\int d^2\bska\frac{(\kappa^{(i)})^2}{D_1D_2} = \pi\bigg[-f_1 \ f_3+\frac{1}{2}+\frac{(q^{(i)})^2}{q^2}\left[\left(1+\frac{2m^2}{q^2}\right)\frac{f_3}{2f_1}-1\right]+\frac{1}{2}\log\bigg(\frac{\Lambda^2}{m^2(1-x)^2}\bigg)\bigg], \tag{A5}\\
			&\int d^2\bska \frac{(\kappa^{(i)})^3}{D_1D_2} = \pi (1-x) \bigg[- \frac{3}{4} q^{(i)} \bigg[ 1+\log{\bigg(\frac{\Lambda^2}{m^2(1-x)^2}\bigg)} \bigg] + \frac{3}{2}\frac{(q^{(i)})^3}{q^2} + \bigg( \frac{6f_1}{4}q^{(i)} - \frac{(q^{(i)})^3}{q^2} \frac{4f_1^2-\frac{m^2}{q^2}}{2f_1} \bigg) f_3 \bigg], \tag{A6}\\
			&\int d^2\bska \frac{(\kappa^{(1)})^2\kappa^{(2)}}{D_1D_2} = \pi (1-x) \bigg[- \frac{1}{4} q^{(2)} \bigg[ 1+\log{\bigg(\frac{\Lambda^2}{m^2(1-x)^2}\bigg)} \bigg] + \frac{3}{2}\frac{(q^{(1)})^2 q^{(2)}}{q^2} \bigg. \\
			& \hspace{9.95cm}+ \bigg( \frac{2f_1}{4}q^{(2)} - \frac{(q^{(1)})^2 q^{(2)}}{q^2} \frac{4f_1^2-\frac{m^2}{q^2}}{2f_1} \bigg) f_3 \bigg], \tag{A7}\\
			&\int d^2\bska \frac{\kappa^{(1)}(\kappa^{(2)})^2}{D_1D_2} = \pi (1-x) \bigg[- \frac{1}{4} q^{(1)} \bigg[ 1+\log{\bigg(\frac{\Lambda^2}{m^2(1-x)^2}\bigg)} \bigg] + \frac{3}{2}\frac{q^{(1)} (q^{(2)})^2}{q^2} \\
			&\hspace{9.95cm}+ \bigg( \frac{2f_1}{4}q^{(1)} - \frac{q^{(1)}( q^{(2)})^2}{q^2} \frac{4f_1^2-\frac{m^2}{q^2}}{2f_1} \bigg) f_3 \bigg], \tag{A8}
		\end{align*}
		where, $i= (1,2)$ and
		\begin{align*}
			D_1&:= \bskasq +m^2(1-x)^2, \tag{A9}\\D_2&:=\left(\bska +(1-x)\bsq\right)^2+m^2(1-x)^2, \tag{A10}\\	f_1 &:=\frac{1}{2}\sqrt{1+\frac{4 m^2}{q^2}}, \tag{A11}\\ f_2&:=\log\left(1+\frac{q^2\left(1+2f_1\right)}{2 m^2}\right), \tag{A12}\\
			f_3&:= \log\left(\frac{1+2 f_1}{-1+2 f_1}\right). \tag{A13}
		\end{align*}

		\section{Steps used to calculate AM distributions}\label{appB}
		
		\subsection{Kinetic Orbital angular momentum}
		\begin{align}
			\langle L^z \rangle (\bsb)=\left. -i\epsilon^{3jk} \int \frac{d^2\boldsymbol{\Delta}^{\perp}}{(2\pi)^2}e^{-i\bsb \cdot \bsD}\frac{\partial \langle T^{+k}_{\text{kin,q}}\rangle_{\text{LF}} }{\partial \Delta_{\perp}^j}\right \vert_{\text{DY}} \label{Lz_kin}.
		\end{align}
		Define, in the Drell-Yan frame:
		\begin{align*}
			\langle T^{\mu \nu} \rangle = \frac{\langle p^{\prime},s|T^{\mu \nu}(0)|p,s \rangle}{2p^+}.
		\end{align*}
		We get from Eq. \ref{T_kin}:
		\begin{align}
			T^{+k}_{\text{kin,q}}(x)=& \frac{1}{2}\overline{\psi}(x)\gamma^{+}i\overleftrightarrow{D}^{k}\psi(x) = \psi_{+}^{\dagger}\left(i\overleftrightarrow{\partial}^{k}+2gA^k\right)\psi_{+},\\ \nonumber T^{+k}_{\text{kin,q}}(0)=& \sum_{\lambda,\lambda^{\prime}}\int \frac{dk^{\prime+}d^2k^{\prime \perp}dk^{+}d^2k^{\perp}}{(16\pi^3)^2\sqrt{k^{\prime+}k^+}}b_{\lambda^{\prime}}^{\dagger}(k^{\prime})b_{\lambda}(k) ~ \chi_{\lambda^{\prime}}^{\dagger}\left[k^{\prime k}+k^{k}\right]\chi_{\lambda}\\&+2g \sum_{\lambda^{\prime},\lambda,\lambda_3}\int \frac{dk^{\prime+}d^2k^{\prime \perp}dk^{+}d^2k^{\perp}dk_3^{+}d^2k_3^{\perp}}{(16\pi^3)^3k_3^{+}\sqrt{k^{\prime+}k^+}}~\chi^{\dagger}_{\lambda^{\prime}}\left[\epsilon_{\lambda_3}^{k}b_{\lambda^{\prime}}^{\dagger}(k^{\prime})b_{\lambda}(k)a_{\lambda_3}(k_3)+\epsilon_{\lambda_3}^{k*}b_{\lambda^{\prime}}^{\dagger}(k^{\prime})b_{\lambda}(k)a^{\dagger}_{\lambda_3}(k_3)\right]\chi_{\lambda}.
		\end{align}
		\textbf{\underline {Diagonal matrix element of $T^{+k}_{\text{kin,q}}(0)$:}}
		\begin{align}
			\nonumber &\frac{\langle 2,\uparrow|T^{+k}_{\text{kin,q}}(0)|2,\uparrow \rangle}{2p^+} \\=& \frac{1}{2} \sum_{\lambda_1^{\prime},\lambda_1, \lambda_2}\int dx d^2\bska ~ \phi^{*\uparrow}_{\lambda_1^{\prime},\lambda_2}\left(x, \bskapr\right) \chi_{\lambda_1^{\prime}}^{\dagger}\left(2\boldsymbol{\kappa}^{k}+\left(1-x\right)\boldsymbol{\Delta}^{k}\right)\chi_{\lambda_1} \phi^{\uparrow}_{\lambda_1,\lambda_2}\left(x,\bska\right),\\\nonumber=& g^2C_F\int \frac{dx d^2\bska}{16 \pi^3}\frac{\left(2\kappa^{k}+\left(1-x\right)\Delta^{k}\right)}{\left(1-x\right)D_1D_2}\times \\& \left[m^2\left(1-x\right)^4+\left(1+x^2\right)\kappa^{\perp2}+\left(1-x\right)\left(1+x^2\right)\bska \cdot \bsD+i\left(1-x\right)\left(1-x^2\right)\left(\kappa^{(1)}\Delta^{(2)}-\kappa^{(2)}\Delta^{(1)}\right)\right],
		\end{align}
		where $x$ is the quark momentum fraction and $\bskapr=\bska+(1-x)\bsD$.\\
		Here we have defined \begin{align}
			D_1 =& \bskasq+m^2\left(1-x\right)^2, \\
			D_2=& \left(\bska+\left(1-x\right)\bsD\right)^2+m^2\left(1-x\right)^2.
		\end{align}
		After performing the $\kappa$ integrations as in Appen. \ref{appA} we get
		\begin{align}
			\nonumber  \frac{\langle 2,\uparrow|T^{+1}_{\text{kin,q}}(0)|2,\uparrow \rangle}{2p^+}  =& i g^2C_F\int \frac{dx}{16 \pi^2}\left(1-x^2\right)\Delta^{(2)}\left[1-\omega~ log\left(\frac{1+\omega}{-1+\omega}\right)+log\left(\frac{\Lambda^2}{m^2\left(1-x\right)^2}\right)\right],  \\=& \frac{i g^2C_F}{16 \pi^2}\Delta^{(2)}\left[\frac{13}{9}-\frac{2}{3}\omega~ log\left(\frac{1+\omega}{-1+\omega}\right)+\frac{2}{3}\log\left(\frac{\Lambda^2}{m^2}\right)\right],\label{T+k_kin1}
		\end{align}
		\begin{align}
			\nonumber \frac{\langle 2,\uparrow|T^{+2}_{\text{kin,q}}(0)|2,\uparrow \rangle}{2p^+}  =& -i g^2C_F\int \frac{dx}{16 \pi^2}\left(1-x^2\right)\Delta^{(1)}\left[1-\omega~ log\left(\frac{1+\omega}{-1+\omega}\right)+log\left(\frac{\Lambda^2}{m^2\left(1-x\right)^2}\right)\right], \\=& \frac{i g^2C_F}{16 \pi^2}\Delta^{(1)}\left[-\frac{13}{9}+\frac{2}{3}\omega~ log\left(\frac{1+\omega}{-1+\omega}\right)-\frac{2}{3}\log\left(\frac{\Lambda^2}{m^2}\right)\right],
			\label{T+k_kin2}
		\end{align}
		where $\omega = \sqrt{1+\frac{4m^2}{\Delta^2}}$.\\
		\textbf{\underline{Non-diagonal matrix element of $T^{+k}_{\text{kin,q}}(0)$:}}
		\begin{align}
			\nonumber &\frac{1}{2p^+}\left[\langle 1,\uparrow|T^{+k}_{\text{kin,q}}(0)|2,\uparrow \rangle + \langle 2,\uparrow|T^{+k}_{\text{kin,q}}(0)|1,\uparrow \rangle\right]  \\=& \frac{g}{\sqrt{16\pi^3}}\sum_{\lambda_1,\lambda_2}\int \frac{dx d^2\bska}{\sqrt{1-x}}~\left[\psi_1^*(P,\sigma^{\prime}) \chi^\dagger_{\sigma'}\epsilon^{k}_{\lambda_2} \chi_{\lambda_1} \phi^{\sigma}_{\lambda_1,\lambda_2}(x, \bska)+ \phi^{*\sigma^{\prime}}_{\lambda_1,\lambda_2}(x,\bska) \chi^\dagger_{\lambda_1} \epsilon_{\lambda_2}^{k*} \chi_{\sigma} \psi_1(P,\sigma)\right].
		\end{align}
		
		\begin{align}
			\nonumber & \langle 1|T^{+1}_{\text{kin,q}}(0)|2 \rangle + \langle 2|T^{+1}_{\text{kin,q}}(0)|1 \rangle \\=& -g^2C_F\int \frac{dxd^2\bska}{8\pi^3}\left(\frac{1+x}{1-x}\right)\frac{\kappa^{(1)}}{D_1} = 0.  &&\text{...Odd integrand in $\kappa$}
		\end{align}
		\begin{align}
			\nonumber  & \langle 1|T^{+2}_{\text{kin,q}}(0)|2 \rangle + \langle 2|T^{+2}_{\text{kin,q}}(0)|1 \rangle \\=& -g^2C_F\int \frac{dxd^2\bska}{8\pi^3}\left(\frac{1+x}{1-x}\right)\frac{\kappa^{(2)}}{D_1} = 0.   &&\text{...Odd integrand in $\kappa$}
		\end{align}
		So, we only have to evaluate the contribution to $\langle T^{+k} \rangle_{\text{LF}}$ from the diagonal term i.e. Eq.\eqref{T+k_kin1} and \eqref{T+k_kin2}. Now, Eq.\eqref{Lz_kin} can be written as:
		\begin{align}
			\langle L^z \rangle (\bsb)= i \int \frac{d^2\boldsymbol{\Delta}^{\perp}}{(2\pi)^2}e^{-i\bsb \cdot \bsD}\left[ \frac{\partial \langle T^{+1}\rangle_{\text{LF}} }{\partial \Delta_{\perp}^{(2)}}-\frac{\partial \langle T^{+2}\rangle_{\text{LF}} }{\partial \Delta_{\perp}^{(1)}}\right]_{\text{DY}}. \label{Lz_full}
		\end{align}
		From Eq.\eqref{Lz_full} we get:
		\begin{align}
			\langle L^z \rangle (\bsb)=\frac{g^2 C_F}{72 \pi^2} \int \frac{d^2\boldsymbol{\Delta}^{\perp}}{(2\pi)^2}e^{-i\bsb \cdot \bsD} \left[-7+\frac{6}{\omega}\left(1+\frac{2m^2}{\Delta^2}\right)log\left(\frac{1+\omega}{-1+\omega}\right)-6~log\left(\frac{\Lambda^2}{m^2}\right)\right],
		\end{align}
		where $\omega = \sqrt{1+\frac{4m^2}{\Delta^2}}$.

		\subsection{Kinetic Spin angular momentum}
		\begin{align}
			\langle S^z \rangle (\bsb)= \frac{1}{2}\epsilon^{3jk}\int \frac{d^2\bsD}{(2\pi)^2}e^{-i \bsD \cdot \bsb} \langle S^{+jk} \rangle_{\text{LF}} \bigg|_{\text{DY}}.
		\end{align}
		We get from Eq. \ref{S_kin}:
		\begin{align}
			S_{q}^{+jk}=\frac{1}{2} \epsilon^{+jk-}\sum_{\lambda,\lambda^{\prime}}\int \frac{dk^{\prime+}d^2\boldsymbol{k}^{\prime \perp}dk^+d^2\boldsymbol{k}^{\prime \perp}}{\left(16 \pi^{3}\right)^2 \sqrt{k^{\prime +}k^+}}b_{\lambda^{\prime}}^{\dagger}(k^{\prime})b_{\lambda}(k) ~ \left(\chi_{\lambda^{\prime}}^{\dagger}\sigma^{(3)}\chi_{\lambda}\right).
		\end{align}
		\textbf{\underline{Diagonal matrix element of $S_q^{+jk}(0)$:}}
		\begin{align}
			&\frac{\langle 2,\uparrow |S_q^{+jk}(0)|2,\uparrow \rangle}{2p^+} \\ 
			&=\frac{1}{4}\epsilon^{+jk-}\sum_{\lambda_1,\lambda_1^{\prime},\lambda_2}\int dx d^2 \bska \phi_{\lambda_1^{\prime},\lambda_2}^{*\uparrow}\left(x,\bskapr\right)\left(\chi^{\dagger}_{\lambda_1^{\prime}}\sigma^{(3)}\chi_{\lambda_1}\right)\phi_{\lambda_1,\lambda_2}^{\uparrow}\left(x,\bska\right),\label{Sl+koverlap} \\
			&=  \frac{g^2C_F}{4}\epsilon^{+jk-}\int \frac{dxd^2\bska}{8\pi^3} \frac{1}{(1-x)D_1D_2}\times \nonumber \\
			& \hspace{0.4cm} \left[ {\kappa^{\perp2}} (1 + x^2) + \bska \cdot \bsD (1 - x)(1 + x^2) + i(1-x)(1 - x^2) (\kappa^{(1)}\Delta^{(2)}-\kappa^{(2)}\Delta^{(1)}) - m^2(1 - x)^4\right]. \label{Sl+k}
		\end{align}
		\textbf{\underline{Non-diagonal matrix element of $S_q^{+jk}(0)$:}}
		\begin{align}
			\langle 1|S_q^{+jk}(0)|2 \rangle = \langle 2|S_q^{+jk}(0)|1 \rangle = 0.
		\end{align}
		After performing the $\kappa$ integration, we get:
		\begin{align}
			\langle S^z \rangle (\bsb) = -&\frac{g^2 C_F}{32 \pi^2} 
			\int \frac{d^2\bsD}{(2\pi)^2}e^{-i \bsD \cdot \bsb}  \int \frac{dx}{1-x} \times \nonumber \\
			&\left[ \omega (1+x^2) \log{\left( \frac{1+\omega}{-1+\omega} \right)} + \left(\frac{1-\omega^2}{\omega}\right) x \log{\left( \frac{1+\omega}{-1+\omega} \right)} - (1+x^2) \log{\left( \frac{\Lambda^2}{m^2(1-x)^2} \right)} \right].
		\end{align}
		\subsection{Belinfante form of total angular momentum}
		
		\begin{align}
			\langle J^z \rangle (\bsb)=\left. -i\epsilon^{3jk} \int \frac{d^2\boldsymbol{\Delta}^{\perp}}{(2\pi)^2}e^{-i\bsb \cdot \bsD}\frac{\partial \langle T^{+k}_{\text{Bel,q}}\rangle_{\text{LF}} }{\partial \Delta_{\perp}^j}\right \vert_{\text{DY}} \label{Jz_kin}.
		\end{align}
		Now, 
		\begin{align}
			T^{+k}_{\text{Bel,q}} =& \frac{1}{4}\overline{\psi}(x)\left[\gamma^{+}i\overleftrightarrow{D}^{k}+\gamma^{k}i\overleftrightarrow{D}^{+}\right]\psi(x),\\=&\underbrace{\frac{1}{2}T^{+k}_{\text{kin,q}}(x)}_{\text{1st term}}+\underbrace{\frac{1}{4}\overline{\psi}(x)\left[\gamma^{k}i \overleftrightarrow{\partial}^{+}\right]\psi(x)}_{\text{2nd term}},
		\end{align}
		where the 1st term has already been calculated. 
		\begin{align}
			&\nonumber \text{2nd term:}\\\nonumber=& \frac{1}{4}\sum_{\lambda,\lambda^{\prime}}\int \frac{dk^{\prime+}d^2\boldsymbol{k}^{\perp}dk^{+}d^2\boldsymbol{k}^{\perp}}{\left(16\pi^3\right)^2\sqrt{k^{\prime+}k^+}}e^{ik^{\prime}\cdot y}\chi_{\lambda^{\prime}}^{\dagger}\left[\sigma^{k}\left(k^{+}+k^{\prime+}\right)\left(\frac{1}{k^{+}}\right)\left(\sigma^{i}k^{i}+im\right)+\left(\sigma^{i}k^{\prime i}-im\right)\left(\frac{1}{k^{\prime+}}\right)\sigma^{k}\left(k^{+}+k^{\prime +}\right)\right]\chi_{\lambda} \\\nonumber & e^{-i k \cdot y} b_{\lambda^{\prime}}^{\dagger}(k^{\prime})b_{\lambda}(k)\\\nonumber&+\frac{g}{4}\sum_{\lambda,\lambda^{\prime},\lambda_{3}}\int \frac{dk^{\prime+}d^2\boldsymbol{k}^{\perp}dk^{+}d^2\boldsymbol{k}^{\perp}dk_{3}^{+}d^2\boldsymbol{k_3}^{\perp}}{\left(16\pi^3\right)^3k_{3}^{+}\sqrt{k^{\prime+}k^+}}e^{ik^{\prime}\cdot y}\chi_{\lambda^{\prime}}^{\dagger}\bigg[\sigma^{k}\bigg\{\frac{k^{+}+k_3^{+}+k^{\prime +}}{k^{+}+k_3^{+}}\left(\sigma^{i}\epsilon_{\lambda_3}^{i}\right)a_{\lambda_3}(k_3)e^{-ik_3 \cdot y}\\\nonumber&+\frac{k^+-k_3^++k^{\prime +}}{k^{+}-k_3^{+}}\left(\sigma^{i}\epsilon_{\lambda_3}^{i*}\right)a^{\dagger}_{\lambda_3}(k_3)e^{ik_3 \cdot y} \bigg\} + \bigg\{\left(\sigma^{i}\epsilon_{\lambda_3}^{i}\right)\frac{k^{+}+k^{\prime +}-k_3^{+}}{k^{\prime +}-k_3^{+}}a_{\lambda_3}(k_3)e^{-ik_3 \cdot y}\\ &+\left(\sigma^{i}\epsilon_{\lambda_3}^{i*}\right)\frac{k^{+}+k^{\prime +}+k_3^+}{k^{\prime +}+k_3^{+}}a^{\dagger}_{\lambda_3}(k_3)e^{ik_3 \cdot y}\bigg\}\sigma^{k}\bigg]\chi_{\lambda}e^{-i k \cdot y}b_{\lambda^{\prime}}^{\dagger}(k^{\prime})b_{\lambda}(k).
		\end{align}
		\textbf{\underline{Diagonal matrix element of $T^{\mu\nu}_{\text{Bel,q}}(0)$:}}
		
		\begin{align}
			&\nonumber \frac{\langle 2,\uparrow|T^{+k}_{\text{Bel,q}}|2,\uparrow \rangle}{2p^+} \\\nonumber  =& \frac{1}{2}\sum_{\lambda_2,\lambda_1,\lambda_1^{\prime}}\int dx d^2\bska~ \phi^{*\sigma^{\prime}}_{\lambda_1^{\prime},\lambda_2}\left(x,\bskapr\right)\chi^{\dagger}_{\lambda_1^{\prime}}\bigg[\left(\boldsymbol{\kappa}^k+\left(1-x\right)\frac{\boldsymbol{\Delta}^k}{2}\right)+\left(\boldsymbol{\sigma}^{i}\left(\boldsymbol{\kappa}^{\prime i}+x\frac{\boldsymbol{\Delta}^{i}}{2}\right)-im\right)\frac{\boldsymbol{\sigma}^{k}}{2}\\&+\frac{\boldsymbol{\sigma}^{k}}{2}\left(\boldsymbol{\sigma}^{i}\left(\boldsymbol{\kappa}^{i}-x\frac{\boldsymbol{\Delta}^{i}}{2}\right)+im\right) \bigg]\chi_{\lambda_1}\phi^{\sigma}_{\lambda_1,\lambda_2}\left(x,\bska\right),
		\end{align}
		where $x$ is the quark momentum fraction and $\bskapr=\bska+(1-x)\bsD$.
		\begin{align}
			&\nonumber \frac{\langle 2,\uparrow|T^{+1}_{\text{Bel,q}}|2,\uparrow \rangle}{2p^+} \\\nonumber=& g^2C_F \int \frac{dxd^2\bska}{32\pi^3}\frac{1}{\left(1-x\right)}\frac{m^2\left(1-x\right)^4\left(4\kappa^{(1)}+2\left(1-x\right)\Delta^{(1)}+i\Delta^{(2)}\right)}{D_1D_2} ~\times\\& \frac{\left(4\kappa^{(1)}+2\left(1-x\right)\Delta^{(1)}-i\Delta^{(2)}\right)\left[\left(1+x^2\right)\left(\kappa^{\perp2}+\left(1-x\right)\bska \cdot \bsD\right)+i\left(1-x\right)\left(1-x^2\right)\left(\kappa^{(1)}\Delta^{(2)}-\kappa^{(2)}\Delta^{(1)}\right)\right]}{D_1D_2}
			\label{T+1_Bel}.
		\end{align}
		\begin{align}
			&\nonumber \frac{\langle 2,\uparrow|T^{+2}_{\text{Bel,q}}|2,\uparrow \rangle}{2p^+} \\\nonumber=& g^2C_F \int \frac{dxd^2\bska}{32\pi^3}\frac{1}{\left(1-x\right)}\frac{m^2\left(1-x\right)^4\left(4\kappa^{(2)}+2\left(1-x\right)\Delta^{(2)}-i\Delta^{(1)}\right)}{D_1D_2} ~\times\\& \frac{\left(4\kappa^{(2)}+2\left(1-x\right)\Delta^{(2)}+i\Delta^{(1)}\right)\left[\left(1+x^2\right)\left(\kappa^{\perp2}+\left(1-x\right)\bska \cdot \bsD\right)+i\left(1-x\right)\left(1-x^2\right)\left(\kappa^{(1)}\Delta^{(2)}-\kappa^{(2)}\Delta^{(1)}\right)\right]}{D_1D_2}.
			\label{T+2_Bel}
		\end{align}
		\textbf{\underline{Non-diagonal matrix element of $T^{\mu\nu}_{\text{Bel,q}}(0)$:}}
		\begin{align}
			&\nonumber \frac{1}{2p^+}\left[\langle 1,\uparrow|T^{+k}_{\text{Bel,q}}|2,\uparrow \rangle + \langle 2,\uparrow|T^{+k}_{\text{Bel,q}}|1,\uparrow \rangle\right] \\ \nonumber =&\frac{g}{4\sqrt{16\pi^3}}\int dx d^2\bska~ \frac{1}{\sqrt{\left(1-x\right)}}\bigg[ \chi^{\dagger}_{\sigma^{\prime}}\left(\boldsymbol{\sigma}^{k}\left(\boldsymbol{\sigma}^{i}\boldsymbol{\epsilon}^{i}_{\lambda_2}\right)+\left(\boldsymbol{\sigma}^{i}\boldsymbol{\epsilon}^{i}_{\lambda_2}\right)\boldsymbol{\sigma}^{k} + 2 \boldsymbol{\epsilon}^{k}_{\lambda_2} \right)\chi_{\lambda_1}\phi_{\lambda_1,\lambda_2}^{\sigma}\left(x,\bska\right)\\ \nonumber&+\phi_{\lambda_1,\lambda_2}^{*\sigma^{\prime}}\left(x,\bska\right)\chi^{\dagger}_{\lambda_1}\left(\left(\boldsymbol{\sigma}^{i}\boldsymbol{\epsilon}^{i*}_{\lambda_2}\right)\boldsymbol{\sigma}^{k}+\boldsymbol{\sigma}^{k}\left(\boldsymbol{\sigma}^{i}\boldsymbol{\epsilon}^{i*}_{\lambda_2}\right) + 2 \boldsymbol{\epsilon}^{k*}_{\lambda_2} \right)\chi_{\sigma}\bigg].
		\end{align}
		This spinor product gives an integrand that is odd in $\kappa$. Thus, after $\kappa$ integration this term vanishes and so we have no contribution from the off-diagonal terms.
		\begin{align}
			\nonumber & \langle 1|T^{+1}_{\text{Bel,q}}(0)|2 \rangle + \langle 2|T^{+1}_{\text{Bel,q}}(0)|1 \rangle \\=& -g^2C_F\int \frac{dxd^2\bska}{8\pi^3}\left(\frac{1+x}{1-x}\right)\frac{\kappa^{(1)}}{D_1} = 0.  &&\text{...Odd integrand}
		\end{align}
		\begin{align}
			\nonumber  & \langle 1|T^{+2}_{\text{Bel,q}}(0)|2 \rangle + \langle 2|T^{+2}_{\text{Bel,q}}(0)|1 \rangle \\=& -g^2C_F\int \frac{dxd^2\bska}{8\pi^3}\left(\frac{1+x}{1-x}\right)\frac{\kappa^{(2)}}{D_1} = 0.   &&\text{...Odd integrand}
		\end{align}
		So, we only have to evaluate the contribution of diagonal terms i.e Eq.\eqref{T+1_Bel} and Eq.\eqref{T+2_Bel} to $T^{+k}_{\text{Bel,q}}$. Belinfante total angular momentum distribution is given as:
		\begin{align}
			\langle J^z \rangle (\bsb)=&\left. -i\epsilon^{3jk} \int \frac{d^2\boldsymbol{\Delta}^{\perp}}{(2\pi)^2}e^{-i\bsb \cdot \bsD}\frac{\partial \langle T^{+k}_{\text{Bel,q}}\rangle_{\text{LF}} }{\partial \Delta_{\perp}^j}\right \vert_{\text{DY}}, \\
			=& i \int \frac{d^2\boldsymbol{\Delta}^{\perp}}{(2\pi)^2}e^{-i\bsb \cdot \bsD}\left[ \frac{\partial \langle T^{+1}_{\text{Bel,q}}\rangle_{\text{LF}} }{\partial \Delta_{\perp}^{(2)}}-\frac{\partial \langle T^{+2}_{\text{Bel,q}}\rangle_{\text{LF}} }{\partial \Delta_{\perp}^{(1)}}\right]_{\text{DY}}.
		\end{align}
		After performing $\kappa$ integration on Eq.\eqref{T+1_Bel} and Eq.\eqref{T+2_Bel} and substituting them in the above equation, we get:
		\begin{align}
			&\nonumber \langle J^z \rangle (\bsb)\\\nonumber=&\frac{g^2 C_F}{16 \pi^2}  \int \frac{d^2\boldsymbol{\Delta}^{\perp}}{(2\pi)^2}e^{-i\bsb \cdot \bsD} \int \frac{dx}{\left(1-x\right)} \frac{1}{{\Delta}^4 \omega^3} ~\times\\\nonumber& \bigg[\left(8m^4\left(1-2x\right)\left(1-x\left(1-x\right)\right)+6m^2\left(1-\left(2-x\right)x\left(1+2x\right)\right)\Delta^2+\left(1-\left(2-x\right)x\left(1+2x\right)\right)\Delta^4\right)\log\left(\frac{1+\omega}{-1+\omega}\right)\\&-\omega \Delta^2 \left(4m^2\left(1-\left(1-x\right)x\right)+\left(1+x^2\right)\Delta^2+\left(1-\left(2-x\right)x\left(1+2x\right)\right)\left(4m^2+\Delta^2\right)log\left(\frac{\Lambda^2}{m^2\left(1-x\right)^2}\right)\right)\bigg].
		\end{align}
		
		\subsection{Correction term}
		\begin{align}
			\langle M_{q}^{z}\rangle(\bsb)=\frac{1}{2}\epsilon^{3jk}\int \frac{d^2\bsD}{(2\pi)^2}e^{-i\bsD \cdot \bsb}\Delta^{l}_{\perp}\frac{\partial\langle S^{l+k}\rangle}{\partial \Delta^{j}_{\perp}}.
		\end{align}
		We substitute from Eq. \ref{Sl+k}: 
		\begin{align}
			\frac{\langle 2,\uparrow |S_q^{l+k}(0)|2,\uparrow \rangle}{2p^+} 
			=&\frac{g^2C_F}{4}\epsilon^{l+k-}\int \frac{dxd^2\bska}{8\pi^3} \frac{1}{(1-x)D_1D_2}\times \nonumber \\
			& \hspace{0.4cm} \left[ {\kappa^{\perp2}} (1 + x^2) + \bska \cdot \bsD (1 - x)(1 + x^2) + i(1-x)(1 - x^2) (\kappa^{(1)}\Delta^{(2)}-\kappa^{(2)}\Delta^{(1)}) - m^2(1 - x)^4\right].
		\end{align}
		After performing the $\kappa$ integration we get
		\begin{align}
			&\nonumber \langle M_{q}^{z}\rangle(\bsb)\\ \nonumber=& -\frac{1}{2}\int \frac{d^2\bsD}{(2\pi)^2}e^{-i\bsD \cdot \bsb}\left(\Delta^{(1)}_{\perp}\frac{\partial}{\partial \Delta^{(1)}}+\Delta^{(2)}_{\perp}\frac{\partial}{\partial \Delta^{(2)}}\right)\left(\frac{\langle 2,\uparrow |S_q^{l+k}(0)|2,\uparrow \rangle}{2p^+}\right)\\\nonumber=& \frac{g^2C_{F}}{32 \pi^2}\int \frac{d^2\bsD}{(2\pi)^2}e^{-i\bsD \cdot \bsb}\int \frac{dx}{\left(1-x\right)}\frac{1}{\omega^3 \Delta^{4}}\\& \left[\omega \Delta^{2}\left(\left(4m^2+\Delta^{2}\right)\left(1+x^2\right)-4m^2x\right)-2m^2\left(\left(4m^2+\Delta^{2}\right)\left(1+x^2\right)-4m^2x-2x\Delta^{2}\right)\right]
		\end{align}
		\bibliographystyle{elsarticle-num}
		\bibliography{references}	
	\end{document}